\documentclass[aps,showpacs,preprintnumbers,amsmath, amssymb]{revtex4}

\usepackage{float}
\usepackage{graphics,epsfig}
\usepackage{graphicx}
\usepackage{dcolumn}
\usepackage{bm}
\usepackage{epstopdf}
\begin{document}
\baselineskip=0.8 cm
\title{{\bf Holographic superconductors in 4D Einstein-Gauss-Bonnet gravity with backreactions}}

\author{Jie Pan$^{1}$, Xiongying Qiao$^{1}$, Dong Wang$^{1}$, Qiyuan Pan$^{1,2}$\footnote{panqiyuan@hunnu.edu.cn}, Zhang-Yu Nie$^{3}$\footnote{niezy@kust.edu.cn}, and Jiliang Jing$^{1,2}$\footnote{jljing@hunnu.edu.cn}}
\affiliation{$^{1}$Key Laboratory of Low Dimensional Quantum Structures and Quantum Control of Ministry of Education, Synergetic Innovation Center for Quantum Effects and Applications, and Department of Physics, Hunan Normal University, Changsha, Hunan
410081, China} \affiliation{$^{2}$Center for Gravitation and Cosmology, College of Physical Science and Technology, Yangzhou University, Yangzhou 225009, China} \affiliation{$^{3}$ Kunming University of Science and Technology, Kunming 650500, China}

\vspace*{0.2cm}
\begin{abstract}
\baselineskip=0.6 cm
\begin{center}
{\bf Abstract}
\end{center}

We construct the holographic superconductors away from the probe limit in the consistent $D\rightarrow4$ Einstein-Gauss-Bonnet gravity. We observe that, both for the ground state and excited states, the critical temperature first decreases then increases as the curvature correction tends towards the Chern-Simons limit in a backreaction dependent fashion. However, the decrease of the backreaction, the increase of the scalar mass, or the increase of the number of nodes will weaken this subtle effect of the curvature correction. Moreover, for the curvature correction approaching the Chern-Simons limit, we find that the gap frequency $\omega_g/T_c$ of the conductivity decreases first and then increases when the backreaction increases in a scalar mass dependent fashion, which is different from the finding in the ($3+1$)-dimensional superconductors that increasing backreaction increases $\omega_g/T_c$ in the full parameter space. The combination of the Gauss-Bonnet gravity and backreaction provides richer physics in the scalar condensates and conductivity in the ($2+1$)-dimensional superconductors.

\end{abstract}

%\keywords{AdS/CFT correspondence, Einstein-Gauss-Bonnet gravity, Holography and condensed matter physics (AdS/CMT)}

\pacs{11.25.Tq, 04.70.Bw, 74.20.-z}\maketitle
\newpage
\vspace*{0.2cm}

\section{Introduction}

As one of the most important developments in the history of theoretical physics, the anti-de Sitter/conformal field theory (AdS/CFT) correspondence \cite{Maldacena,Witten,Gubser1998} has been becoming a powerful tool to study the strongly correlated condensed matter physics in the past decade \cite{JZaanenSLS}. Especially, the ``hair''/``no hair'' transition of the AdS black hole can be used to investigate the corresponding superconductor and its phase transition in the condensed matter systems \cite{GubserPRD78}. By coupling AdS gravity to a Maxwell field and charged scalar, Hartnoll, Herzog and Horowitz constructed the first holographic superconductor model in the probe limit where the backreaction of matter fields on the spacetime metric is neglected, and reproduced the properties of a ($2+1$)-dimensional superconductor \cite{HartnollPRL101}. This pioneering work on this topic has led to many investigations concerning the s-, p- and d-wave holographic superconductor models, which might shed some light on the understanding of the microscopic origins of strongly correlated superconductivity \cite{HartnollJHEP12,FadafanRE}; for reviews, see Refs. \cite{CaiRev,HartnollRev,HerzogRev,HorowitzRev} and references therein.

In most cases, the studies on the holographic superconductors focus on the Einstein-Maxwell theory coupled to a charged (scalar, vector or massive spin two) field. Motivated by the application of the Mermin-Wagner theorem to the holographic dual models, it is of great interest to generalize the investigation on the holographic superconductor model to the Gauss-Bonnet gravity \cite{Cai-2002} and analyze the effect of the curvature correction on the superconductor phase transition, since this will help us to understand the influences of the $1/N$ or $1/\lambda$ ($\lambda$ is the 't Hooft coupling) corrections on the holographic models \cite{LiuFLWZ}. In 2009, Gregory, Kanno and Soda introduced holographic superconductors in the five-dimensional Einstein-Gauss-Bonnet gravity in the probe limit \cite{Gregory}, which shows that the higher curvature corrections make condensation harder and cause the universal behavior of the conductivity $\omega_g/T_c\approx8$ \cite{HorowitzPRD78} unstable. Taking the backreaction of the matter fields into account in the ($3+1$)-dimensional Gauss-Bonnet superconductors, it was found that the critical temperature first decreases then increases as the Gauss-Bonnet term tends towards the Chern-Simons value in a backreaction dependent fashion but the effect of both backreaction and higher curvature is to increase the gap ratio $\omega_g/T_c$ \cite{BarclayGregory,Brihaye}. Further investigations on the holographic dual models based on the Einstein-Gauss-Bonnet gravity in dimensions $D\geq5$ have been carried out, including the s-wave \cite{Pan-Wang,Gregory2011,Ge-Wang,KannoGB,
Gangopadhyay2012,GhoraiGangopadhyay,SheykhiSalahiMontakhab,SalahiSheykhiMontakhab,LiFuNie,CHNam,ParaiEPJC2020}, p-wave \cite{CaiPWaveGB,LiCaiZhang,LuWuNPB2016,GBSuperfluid,MohammadiEPJC2019,LaiEPJC2020}, s$+$p \cite{NieZeng} models and color superconductivity in quantum chromodynamics (QCD) \cite{FadafanRojas}. However, since it was believed that the Gauss-Bonnet term does not contribute to the gravitational dynamics in four dimensions, the study on the ($2+1$)-dimensional Gauss-Bonnet superconductors is called for.

More recently, Glavan and Lin presented a novel four-dimensional ($4D$) Einstein-Gauss-Bonnet gravity by rescaling the Gauss-Bonnet coupling constant $\alpha\rightarrow\alpha/(D-4)$ and taking the limit $D\rightarrow4$, where the Gauss-Bonnet term makes an important contribution to the gravitational dynamics \cite{GlavanLin}. Considering the comments and criticisms on the original version of the 4D Einstein-Gauss-Bonnet gravity \cite{Ai2020,Mahapatra2020,Shu09339,TianZhu,ArrecheaDJ,GursesST}, some researchers proposed the ``regularized" versions of the 4D Einstein-Gauss-Bonnet gravity \cite{LuPang,HennigarKMP,Fernandes08362,OikonomouF} and the consistent theory of the $D\rightarrow4$ Einstein-Gauss-Bonnet gravity \cite{AGM}. In particular, using the consistent $D\rightarrow4$ Einstein-Gauss-Bonnet gravity, the authors of \cite{qiao} obtained the 4D Gauss-Bonnet-AdS black hole solution, which was proved out to be the solution obtained via the scalar-tensor theories with the Gauss-Bonnet term \cite{LuPang,HennigarKMP} as well as the naive $D\rightarrow4$ limit of the higher-dimensional theory \cite{Fernandes,WeiL14275,KonoplyaZhidenko}. Thus, the ($2+1$)-dimensional Gauss-Bonnet superconductors in the probe limit were constructed in Ref. \cite{qiao}, which shows that, different from the finding in the higher-dimensional superconductors that the higher curvature correction makes the scalar hair more difficult to be developed in the full parameter space, the curvature correction has a more subtle effect on the scalar condensates in the s-wave superconductor in ($2+1$)-dimensions, i.e., the critical temperature first decreases then increases as the Gauss-Bonnet parameter tends towards the Chern-Simons value in a scalar mass dependent fashion. For the conductivity in the probe limit, the higher curvature correction results in the larger deviation from the expected relation in the gap frequency $\omega_g/T_c\approx 8$ in ($2+1$)-dimensional models \cite{qiao}. Although the probe approximation is known to capture the essential features of the problem, it would be of great interest to extend the study to take into consideration of the backreaction since it can provide richer physics in the phase transition of the holographic dual models \cite{HartnollJHEP12}. So in this work we will build the holographic superconductors in 4D Einstein-Gauss-Bonnet gravity with the backreactions and examine the influences of the curvature correction and the backreaction on the ($2+1$)-dimensional superconductors. Considering the increasing interest in study of the excited states by holography \cite{WangJHEP2020,QiaoEHS,Liran,XiangZW,OuYangliang,WangLLZEPJC,ZhangZPJ}, we also investigate the excited states of the ($2+1$)-dimensional Gauss-Bonnet superconductors away from the probe limit, which exhibits some interesting and different features when compared to the the ground state.

The structure of this work is as follows. In Sec. II we will construct the ($2+1$)-dimensional Gauss-Bonnet superconductors with the backreactions. In particular, we derive the equations of motion and the boundary conditions for the superconductor model in the consistent $D\rightarrow4$ Einstein-Gauss-Bonnet gravity. In Sec. III we will investigate the effects of the curvature correction and the backreaction on the superconductor phase transition both for the ground state and excited states. In Sec. IV we will explore the effects of the curvature correction and the backreaction on the conductivity of the system. We will conclude in the last section with our main results.

\section{Description of the holographic dual system}

In order to construct the backreacting holographic superconductor in the $4D$ Einstein-Gauss-Bonnet gravity, we start with the Gauss-Bonnet-AdS black hole solution by using the consistent $D\rightarrow4$ Einstein-Gauss-Bonnet gravity \cite{AGM}. In the ADM formalism, we can take the metric ansatz
\begin{equation}\label{metric}
ds^2=g_{\mu\nu} dx^{\mu} dx^{\nu}=-N^{2}dt^{2}+\gamma_{ij}(dx^{i}+N^{i}dt)(dx^{j}+N^{j}dt),
\end{equation}
with the lapse function $N$, spatial metric $\gamma_{ij}$ and shift vector $N^i$. We consider the action containing a $U(1)$ gauge field and the scalar field coupled via a generalized Lagrangian
\begin{eqnarray}\label{4DEGBActionR}
S&=& \int dt d^3x N\sqrt{\gamma}\left(\mathcal{L}^{\rm 4D}_{\rm EGB}-\frac{1}{4}F_{\mu \nu}F^{\mu \nu}-|\nabla \psi-i q A \psi|^{2}-m^{2}|\psi|^{2}\right),
\end{eqnarray}
with the Lagrangian density
\begin{eqnarray}\label{LD-EGBR}
\mathcal{L}^{\rm 4D}_{\rm EGB} &=& \frac{1}{2\kappa^{2}}
\left \{2R+\frac{6}{L^{2}}- \mathcal{M} + \frac{\alpha}{2}
\left [8R^2 -4 R\mathcal{M} -\mathcal{M}^2
- \frac{8}{3} \left(8R_{ij}R^{ij}-4R_{ij}\mathcal{M}^{ij}
-\mathcal{M}_{ij}\mathcal{M}^{ij}\right) \right ] \right \}\,,
\end{eqnarray}
where $\kappa^{2}=8\pi G$ is the gravitational coupling constant and $\alpha$ is the Gauss-Bonnet coupling. $R$ and $R_{ij}$ represent the Ricci scalar and Ricci tensor of the spatial metric, and $q$ and $m$ are the charge and mass of the scalar field $\psi$, and
\begin{eqnarray}\label{MR}
\mathcal{M}_{ij} \equiv R_{ij}+\mathcal{K}^k{}_{k}
\mathcal{K}_{ij}-\mathcal{K}_{ik}\mathcal{K}^k{}_{j},
\hspace{1cm} \mathcal{M} \equiv \mathcal{M}^i{}_{i} \,,
\end{eqnarray}
with $\mathcal{K}_{ij} \equiv  \left [\dot{\gamma}_{ij}-2D_{(i}N_{j)}-\gamma_{ij}D^2 \lambda_{\rm
GF} \right ]/(2N)$. Here, a dot denotes differentiation in the time $t$ and $D_{i}$ is the covariant derivative
compatible with the spatial metric.

For the four-dimensional planar black hole, we adopt the ansatz for the metric
\begin{equation}\label{metric-BGR}
N=\sqrt{f(r)}e^{-\chi(r)/2} \,, \hspace{1cm} N^i = 0 \,, \hspace{1cm} \gamma_{ij}=\text{diag}\left(\frac{1}{f(r)} ,r^2,r^2\right),
\end{equation}
and for the matter fields
\begin{equation}\label{Matterfields}
\psi=\psi(r), \qquad  A_{\mu}=(\phi(r),0,0,0),
\end{equation}
where $\psi(r)$ and $\phi(r)$ are both real functions of $r$ only. Therefore, considering that the Lagrange multiplier $\lambda_{\rm GF}$ can be set to zero for the symmetric static backgrounds \cite{AokiGMJCAP,AokiGMM}, from the action (\ref{4DEGBActionR}) we obtain the equations of motion
\begin{eqnarray}\label{psi}
\psi''+\left(\frac{2}{r}-\frac{\chi'}{2}+\frac{f'}{f}\right)\psi'
+\left(\frac{q^{2}e^{\chi}\phi^2}{f^2}-\frac{m^2}{f}\right)\psi=0,
\end{eqnarray}
\begin{eqnarray}\label{phi}
\phi''+\left(\frac{2}{r}+\frac{\chi'}{2}\right)\phi'-\frac{2q^{2}\psi^2}{f}\phi=0,
\end{eqnarray}
\begin{eqnarray}\label{metricf}
f'-\frac{1}{r^2-2\alpha f}\left(\frac{3r^3}{L^2}-rf-\frac{\alpha f^{2}}{r}\right)+\frac{\kappa^2r^3}{r^2-2\alpha f}\left[m^2\psi^2+\frac{1}{2}e^{\chi}\phi'^2\right.
\left.+f\left(\psi'^2+\frac{q^{2}e^{\chi}\phi^2\psi^2}{f^2}\right)\right]=0,
\end{eqnarray}
\begin{eqnarray}\label{chi}
\chi'+\frac{2\kappa^2r^3}{r^2-2\alpha f}\left(\psi'^2+\frac{q^{2}e^{\chi}\phi^2\psi^2}{f^2}\right)=0,
\end{eqnarray}
where the prime denotes the derivative with respect to $r$. When the Gauss-Bonnet parameter $\alpha\rightarrow0$, Eqs. (\ref{psi})-(\ref{chi}) reduces to Eqs. (2.5)-(2.8) in the standard holographic superconductors with the backreactions for $d=4$ investigated in \cite{PanJWC}. It should be noted that the corresponding Hawking temperature is given by
\begin{eqnarray}\label{HawkingT}
T_{H}=\frac{f'(r_{+})e^{-\chi(r_{+})/2}}{4\pi},
\label{HawkingT}
\end{eqnarray}
which will be interpreted as the temperature of the CFT.

For the superconducting phase, $\psi(r)\neq0$, we should impose the appropriate boundary conditions to solve Eqs. (\ref{psi})-(\ref{chi}) numerically. At the horizon $r=r_{+}$, we have the boundary conditions by requiring that the scalar field $\psi$ and metric coefficient $\chi$ are regular, and the gauge field $\phi$ and metric coefficient $f$ satisfy $\phi(r_{+})=0$ and $f(r_{+})=0$, respectively. Near the asymptotic boundary $r\rightarrow\infty$, we obtain the asymptotic behaviors
\begin{eqnarray}
\chi\rightarrow0\,,\hspace{0.5cm}
f\sim\frac{r^{2}}{L^2_{\rm eff}}\,,\hspace{0.5cm}
\phi\sim\mu-\frac{\rho}{r}\,,\hspace{0.5cm}
\psi\sim\frac{\psi_{-}}{r^{\Delta_{-}}}+\frac{\psi_{+}}{r^{\Delta_{+}}}\,,
\label{infinity}
\end{eqnarray}
with the effective asymptotic AdS scale \cite{Cai-2002}
\begin{eqnarray}\label{LeffAdS}
L^2_{\rm eff}=\frac{2\alpha}{1-\sqrt{1-\frac{4\alpha}{L^2}}},
\end{eqnarray}
where $\Delta_{\pm}=\left(3\pm\sqrt{9+4m^{2}L^2_{\rm eff}}\right)/2$ are the characteristic exponents, and $\mu$ and $\rho$ are interpreted as the chemical potential and charge density in the dual field theory, respectively. In Ref. \cite{qiao}, the authors showed that it is more appropriate to fix the mass of the field by choosing the value of $m^{2}L_{\rm eff}^{2}$, since this choice can disclose the correct consistent influence due to the Gauss-Bonnet parameter in various condensates for all dimensions. Thus, we will choose the mass of the scalar field by selecting values of $m^{2}L_{\rm eff}^{2}$ in this work. On the other hand, considering the so-called Chern-Simons limit $\alpha=L^2/4$  \cite{CrisostomoTZ}, i.e., the upper bound of the Gauss-Bonnet parameter, we will take the range $-L^2/10\leq\alpha\leq L^2/4$ for the Gauss-Bonnet coupling. For simplicity, we scale $L=1$ in the following calculation.

For the normal phase, $\psi(r)=0$, the metric coefficient $\chi$ is a constant. Thus, we get the analytical solutions to Eqs. (\ref{phi}) and (\ref{metricf}), i.e.,
\begin{eqnarray}
f(r)=\frac{r^2}{2\alpha}\left[1-\sqrt{1-\frac{4\alpha}{L^2}\left(1-\frac{r_+^3}{r^3}\right)
+\frac{2\alpha\kappa^2\rho^2}{r_{+}r^3}\left(1-\frac{r_+}{r}\right)}\right], \qquad \phi(r)=\mu-\frac{\rho}{r},
\end{eqnarray}
which are just the four-dimensional charged Gauss-Bonnet black holes in AdS space \cite{WeiL14275}. If $\alpha\rightarrow0$, we can recover the four-dimensional AdS Reissner-Nordstr\"{o}m black hole.

Interestingly, from Eqs. (\ref{psi})-(\ref{chi}) we can obtain the useful scaling symmetries and the transformation of the relevant quantities
\begin{align}
\begin{split}
& r\rightarrow\lambda r,\qquad (t,x,y)\rightarrow\lambda^{-1}(t,x,y),\qquad
\psi\rightarrow\psi ,\qquad  \phi\rightarrow\lambda\phi  ,\qquad q\rightarrow q,\\
&(f,\rho)\rightarrow\lambda^{2}(f,\rho), \qquad (T,\mu)\rightarrow\lambda (T,\mu),
\qquad \psi_{\pm}\rightarrow\lambda^{\Delta_{\pm}}\psi_{\pm},
\end{split}
\end{align}
where $\lambda$ is a real positive number. So we will take advantage of these qualities to set $r_+=1$, $q=1$ and keep $\kappa^{2}$ finite when including the backreaction \cite{PanJWC}.

\section{Condensates of the scalar field}

In Ref. \cite{qiao}, the authors constructed the holographic superconductors in the $4D$ Einstein-Gauss-Bonnet gravity in the probe limit and found that the curvature correction has a more subtle effect on the scalar condensates of the s-wave superconductor in ($2+1$)-dimensions, which is different from the finding in the higher-dimensional superconductors that the higher curvature correction makes the scalar hair more difficult to be developed in the full parameter space. Now we will further investigate the effect of the higher curvature correction on the condensates of the scalar field in the ($2+1$)-dimensional superconductors away from the probe limit.

According to the AdS/CFT correspondence, provided $\Delta_{-}$ is larger than the unitarity bound in Eq. (\ref{infinity}), both $\psi_{-}$ and $\psi_{+}$ can be normalizable and be used to define operators on the dual field theory, $\psi_{-}=\langle{\cal O}_{-}\rangle$, $\psi_{+}=\langle{\cal O}_{+}\rangle$, respectively \cite{HartnollPRL101}. Thus, in the following we will impose boundary condition that either $\psi_{+}$ or $\psi_{-}$ vanishes, just as in \cite{qiao}.

\subsection{Operator ${\cal O}_{+}$}

\begin{figure}[ht]
\includegraphics[scale=0.65]{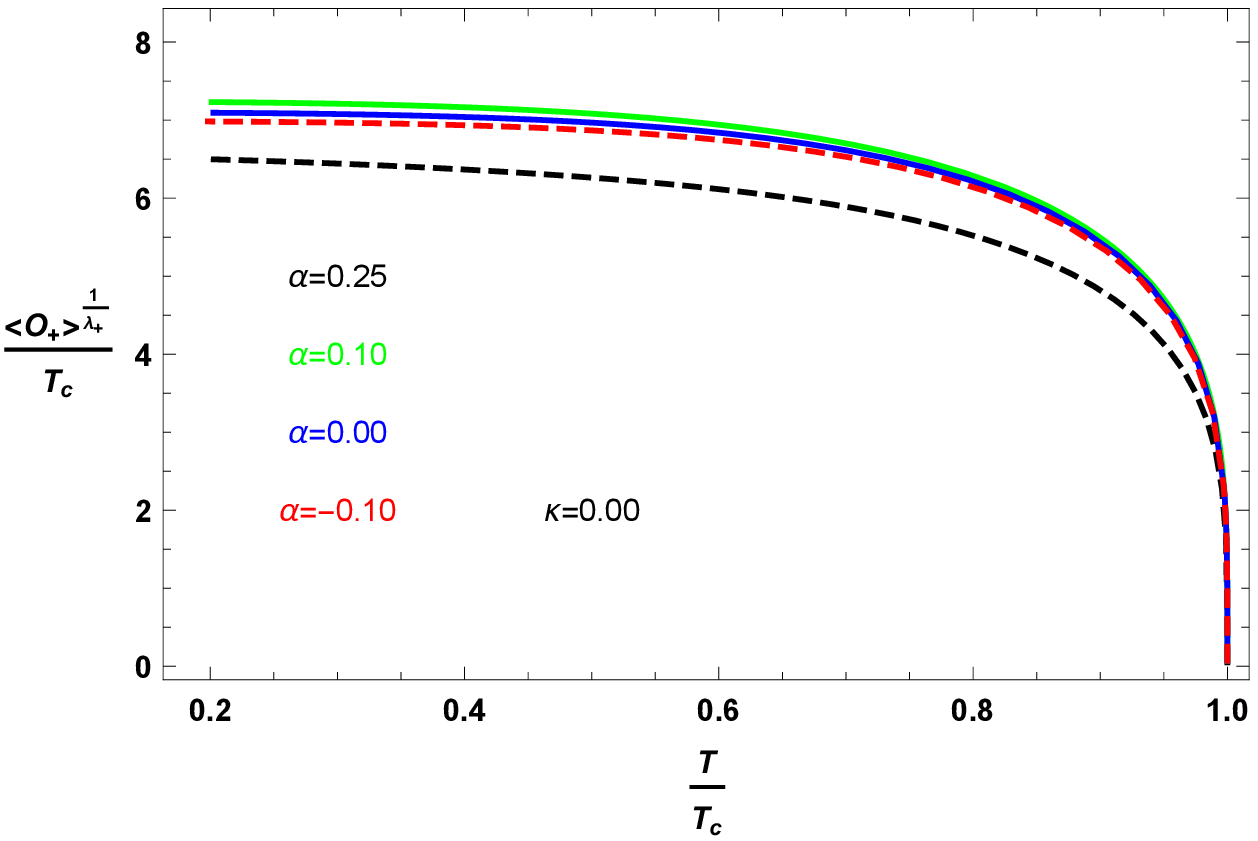}
\includegraphics[scale=0.65]{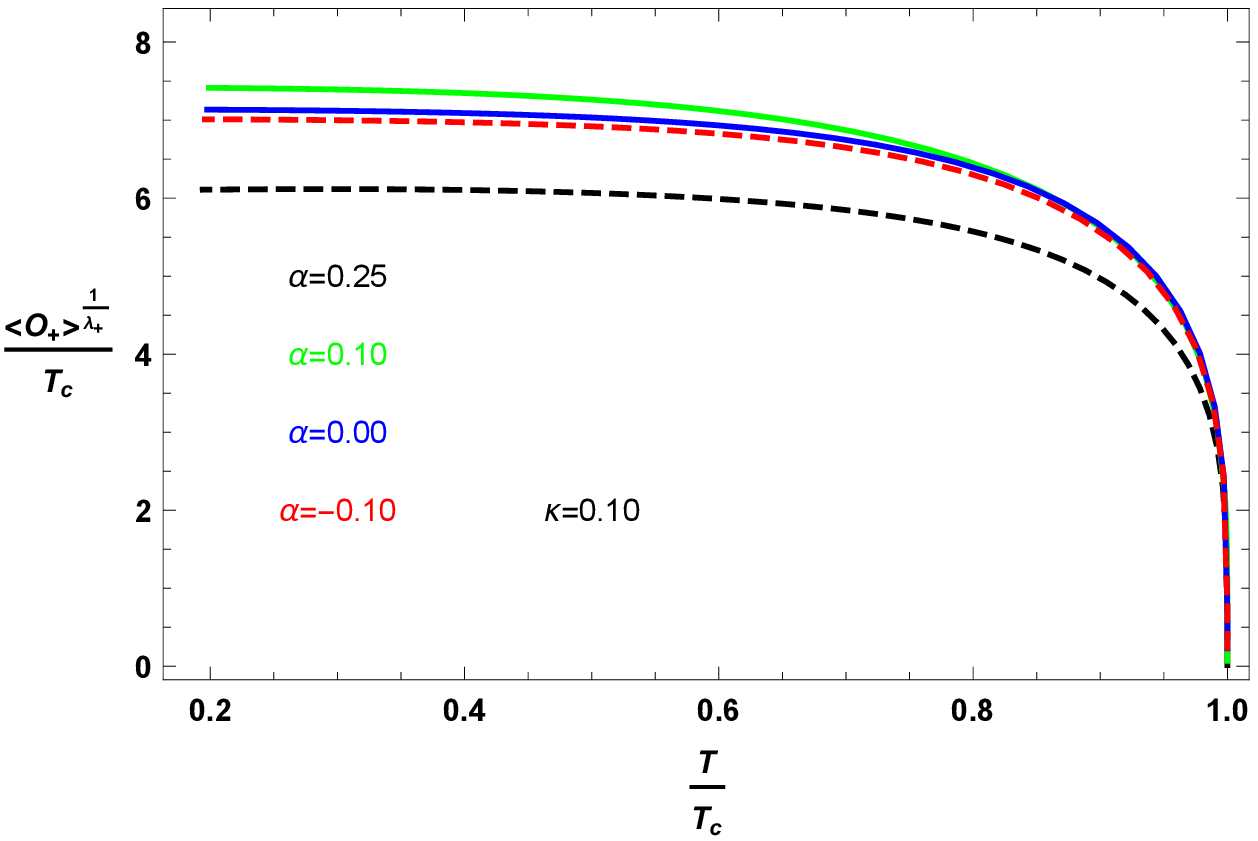}\\
\includegraphics[scale=0.65]{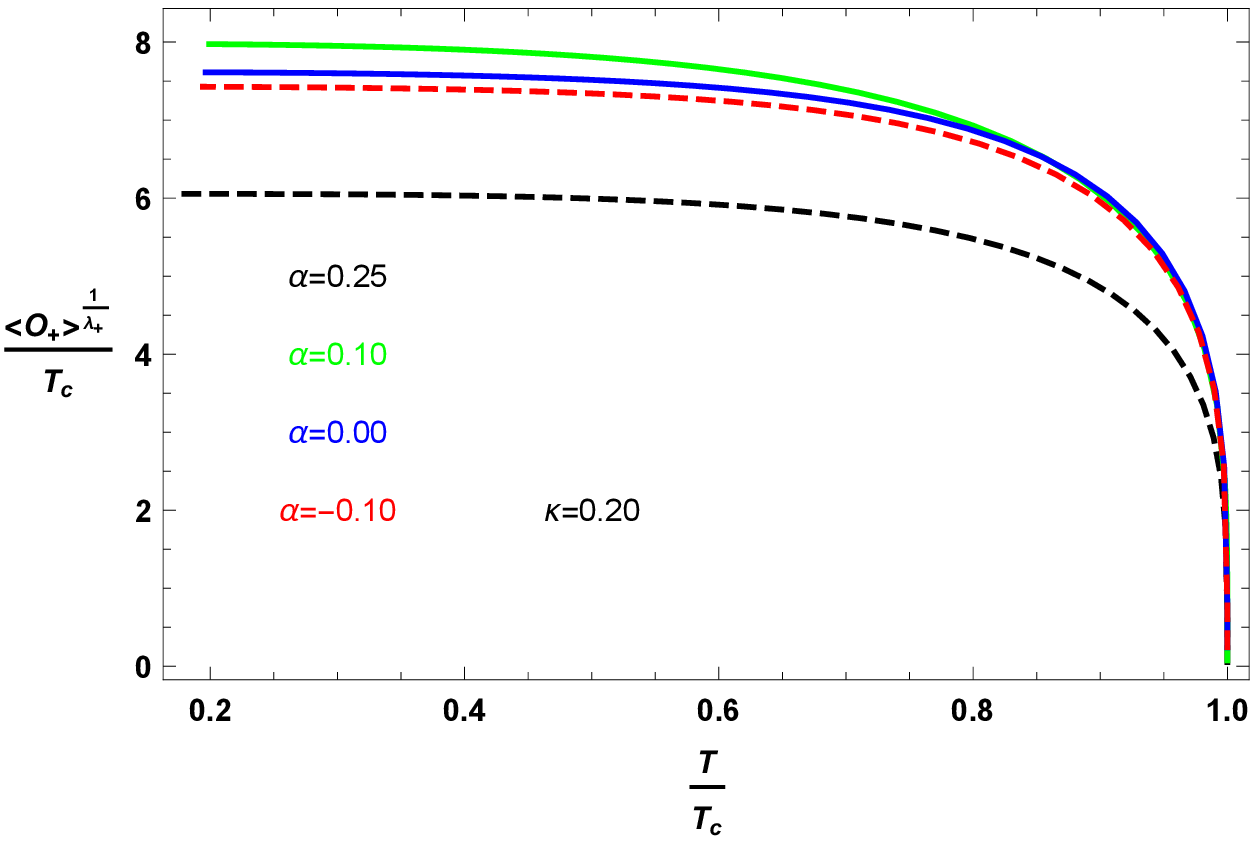}
\includegraphics[scale=0.65]{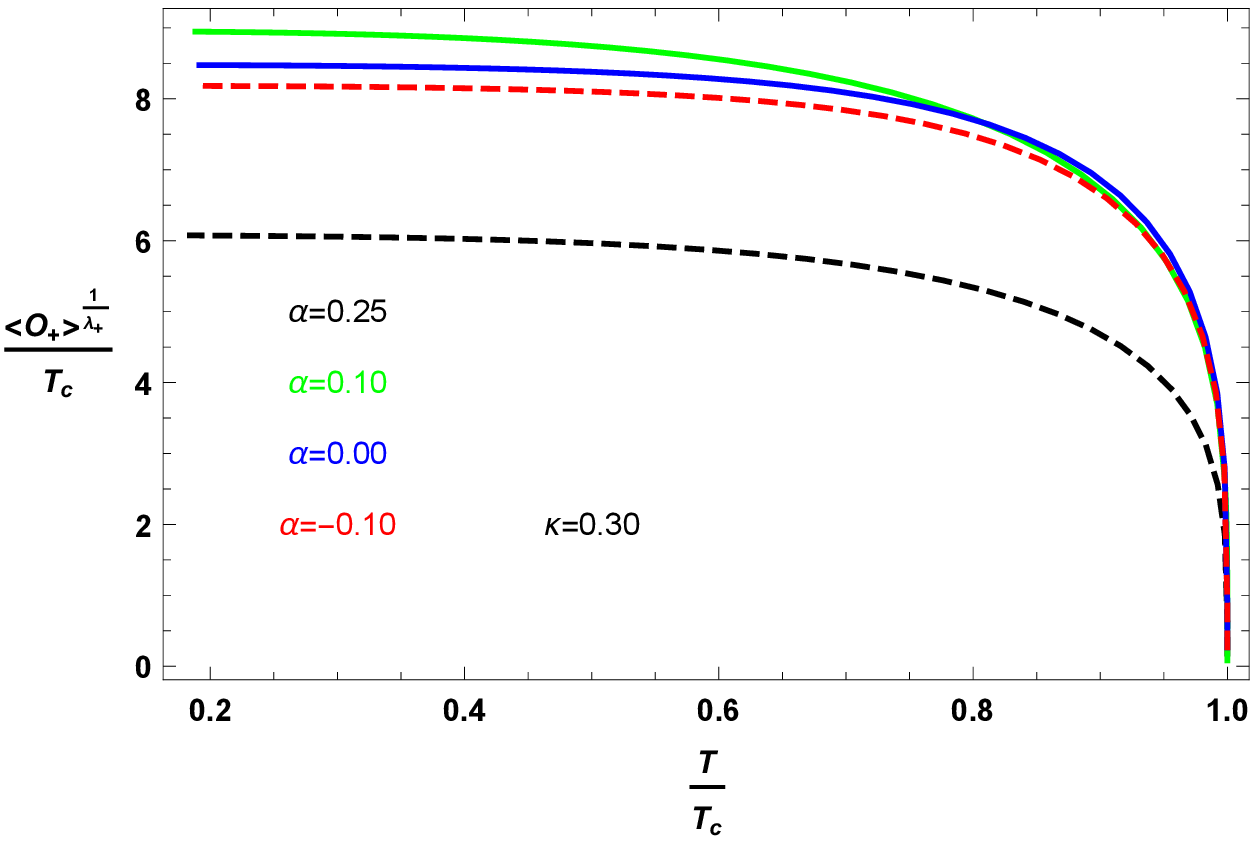}
\caption{\label{figure1} (color online) The condensates of the scalar operator ${\cal O}_{+}$ as a function of temperature for the fixed mass of the scalar field $m^{2}L_{\rm eff}^{2}=-2$ with different Gauss-Bonnet parameters $\alpha$ and backreaction parameters $\kappa$ in the ground state. In each panel, the four lines correspond to the increasing Gauss-Bonnet parameter, i.e., $\alpha=-0.10$ (red and dashed), $0.00$ (blue), $0.10$ (green) and $0.25$ (black and dashed), respectively. }
\end{figure}

In this section, we impose the boundary condition $\psi_{-}=0$ and concentrate on the condensate for the operator ${\cal O}_{+}$. Just as in the standard holographic superconductors \cite{HartnollJHEP12}, we first investigate the ground state, which is the first state to condense \cite{WangSPJ}. In Fig. \ref{figure1}, we exhibit the condensates of the scalar operator ${\cal O}_{+}$ as a function of temperature with various Gauss-Bonnet parameters and backreaction parameters for the fixed mass of the scalar field  $m^2L_{eff}^2=-2$ in the ground state. It is found that, similar to the BCS theory, at zero temperature the condensate goes to a constant which depends on the Gauss-Bonnet parameter $\alpha$ and backreaction parameter $\kappa$. This implies that the s-wave holographic superconductors still exist even we consider Gauss-Bonnet correction terms to the standard $(2+1)$-dimensional holographic superconductor model with the backreactions \cite{HartnollJHEP12,PanJWC}. For small condensate, we can fit these curves and obtain a square root behavior $\langle {\cal O}_{+}\rangle\sim (1-T/T_{c})^{1/2}$, which is shown clearly that, for the ground state, the phase transition of the $4D$ Gauss-Bonnet holographic superconductors with the backreactions belongs to the second order and the critical exponent of the system takes the mean field value $1/2$ for all values of $\alpha$.

\begin{figure}[ht]
\includegraphics[scale=0.65]{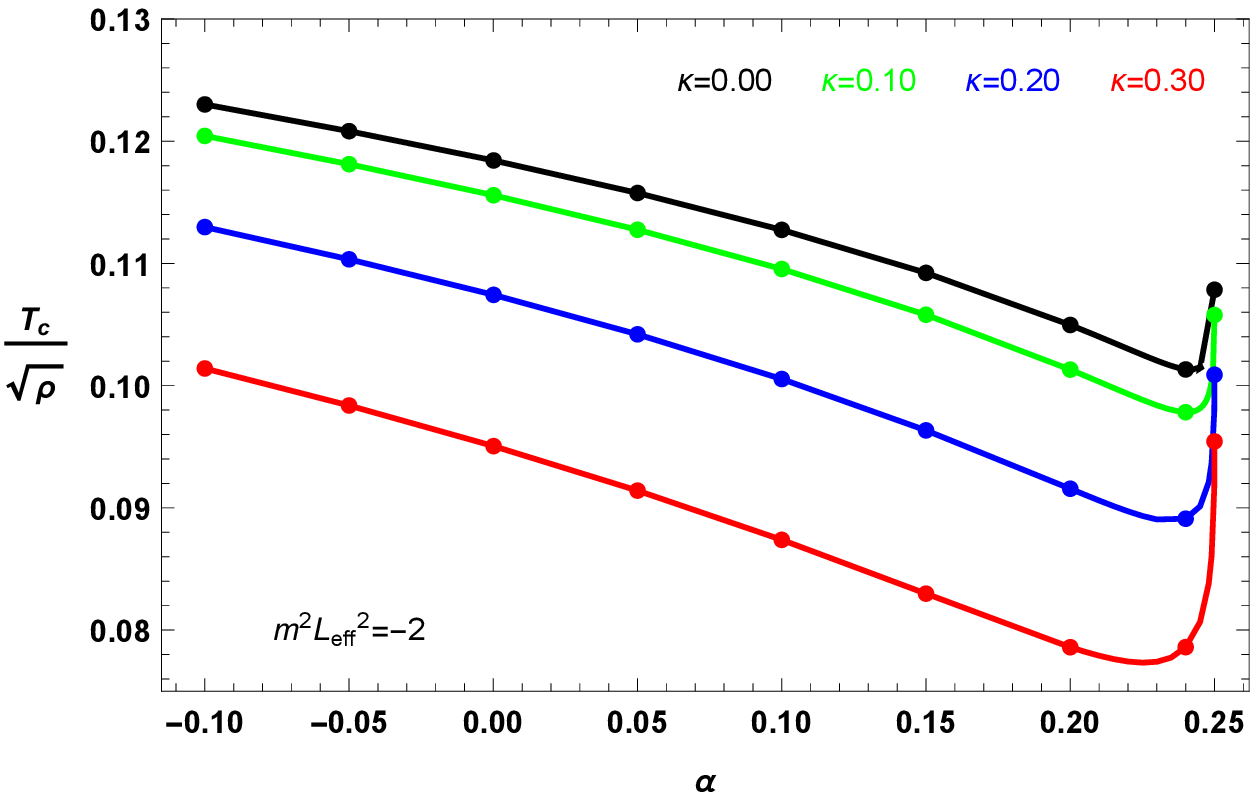}
\includegraphics[scale=0.65]{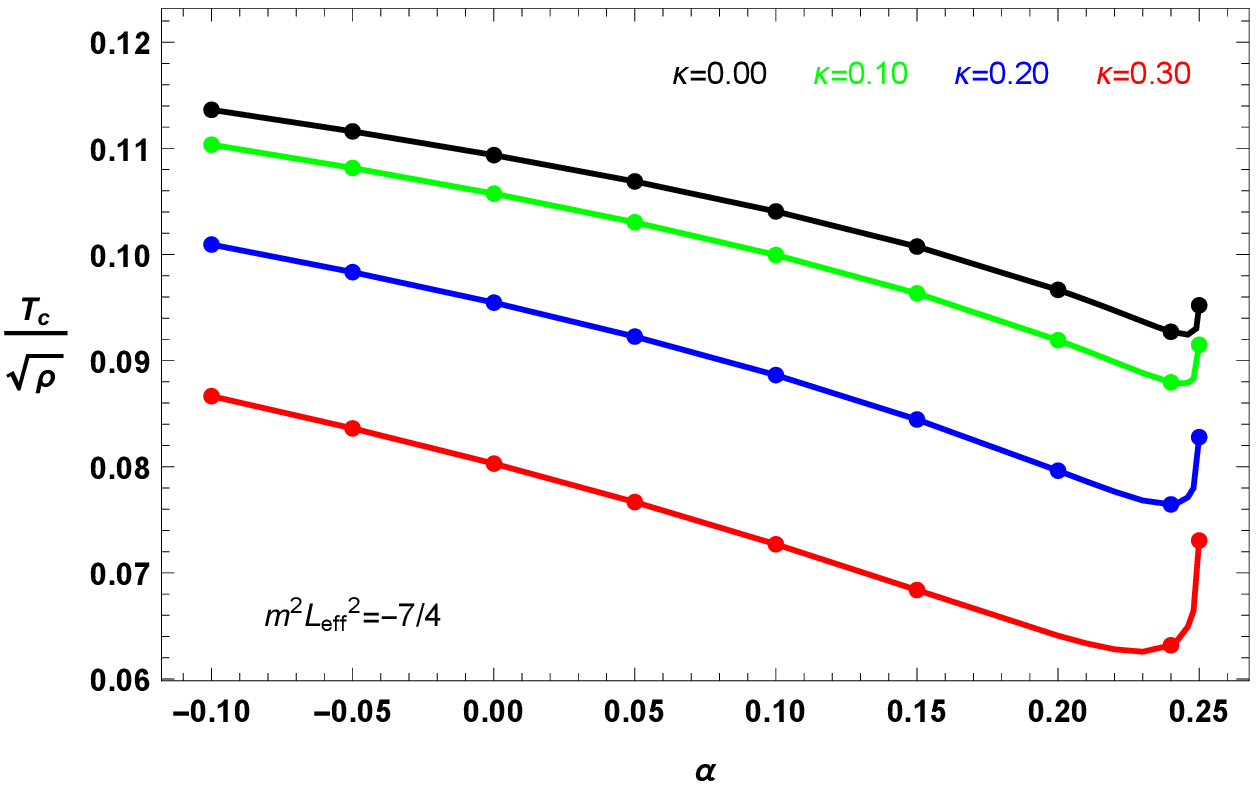}\\
\includegraphics[scale=0.57]{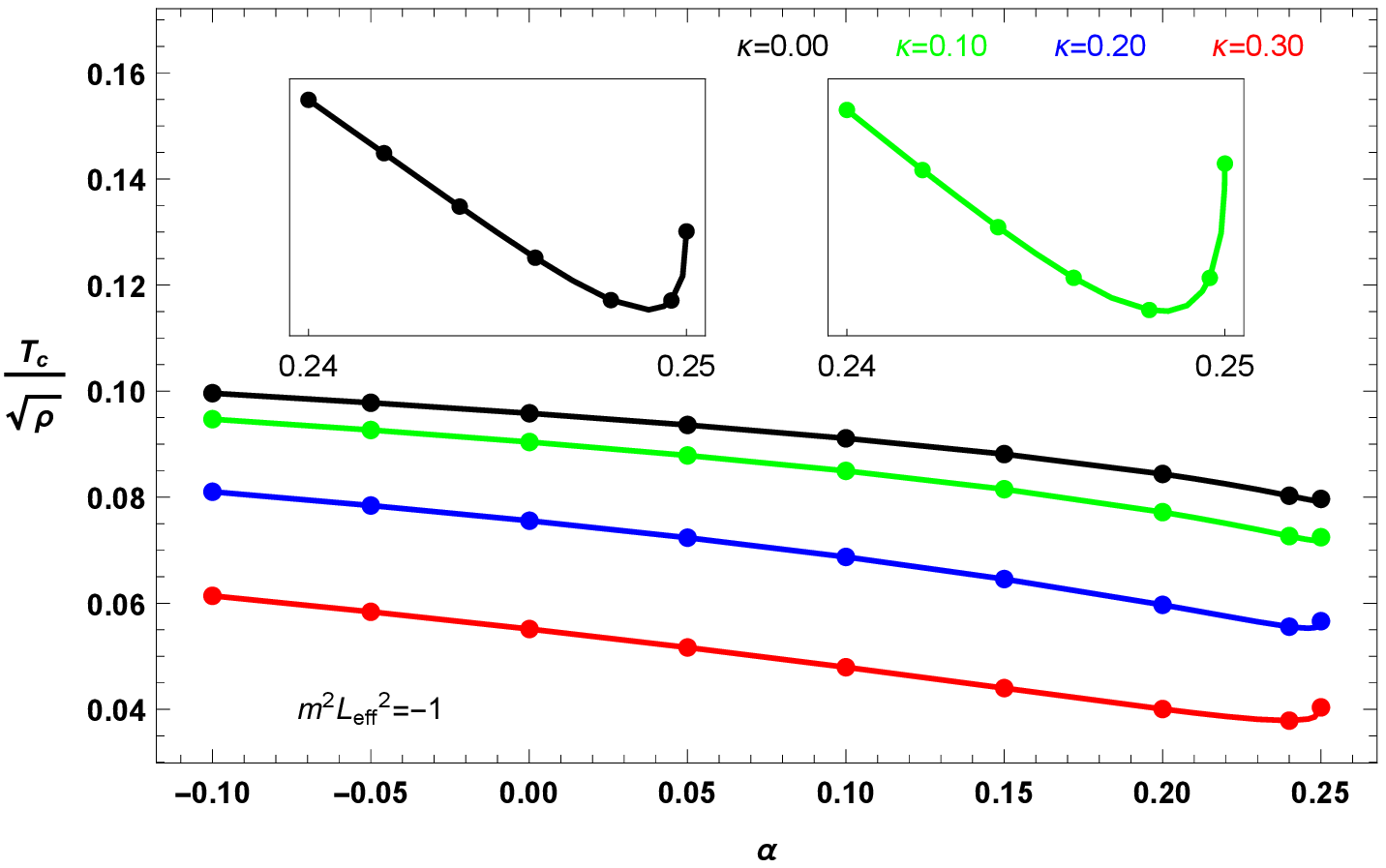}
\includegraphics[scale=0.65]{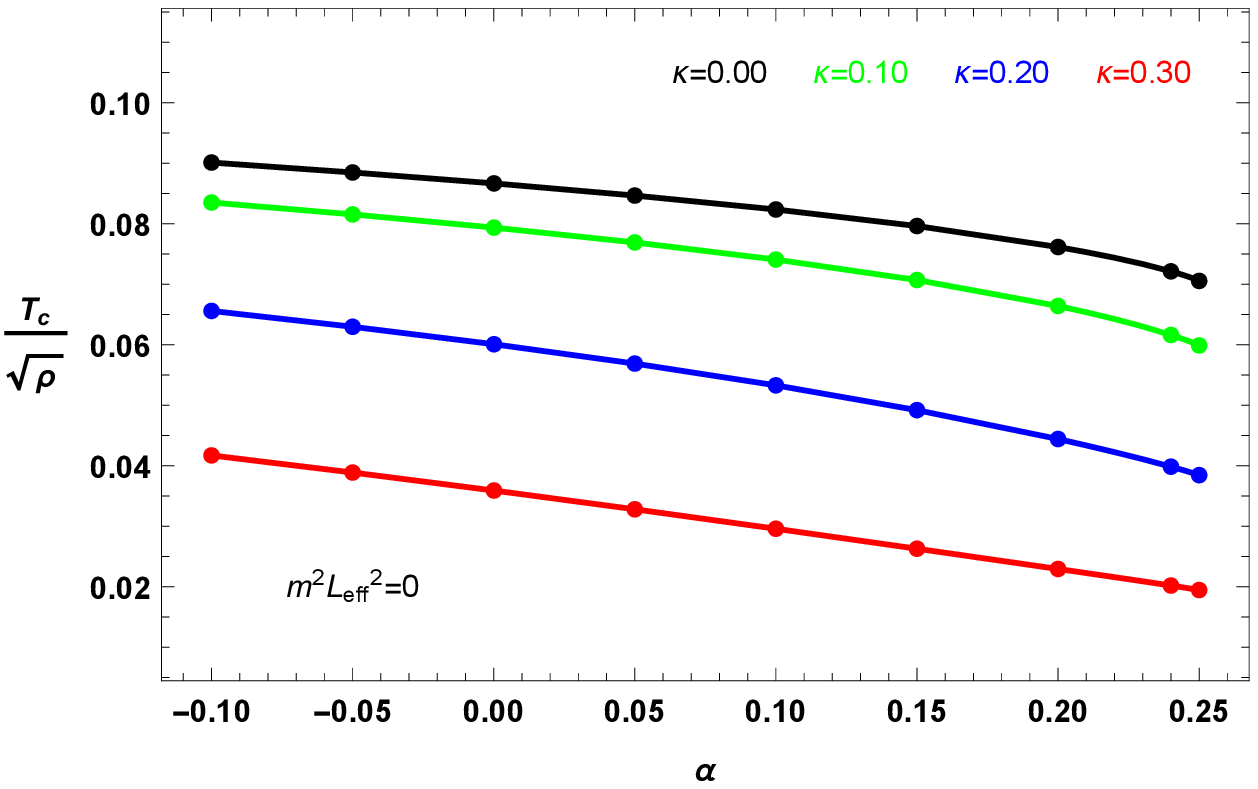}\\
\caption{\label{figure2}(color online) The critical temperature $T_{c}$ of the scalar operator ${\cal O}_{+}$ as a function of the Gauss-Bonnet parameter for the fixed masses of the scalar field with different backreaction parameters $\kappa$ in the ground state. In each panel, the four lines from top to bottom correspond to the increasing backreaction parameters, i.e., $\kappa=0.00$ (black), $0.10$ (green), $0.20$ (blue) and $0.30$ (red), respectively. }
\end{figure}

In order to investigate the effect of the curvature correction on the critical temperature in the ground state, we present the critical temperature $T_c$ as a function of the Gauss-Bonnet parameter for the different choices of the mass of the scalar field and different backreaction parameters $\kappa$ in Fig. \ref{figure2}. For the fixed $m^{2}L_{\rm eff}^{2}$ and $\alpha$, it is easy to see that the effect of the backreaction is to decrease $T_c$, which indicates that the backreaction can hinder the condensate to be formed. Moreover, we observed that, for the small mass scale such as $m^2L_{\rm eff}^2=-2$, $-7/4$ and $-1$, the critical temperature $T_c$ decreases first and then increases with the increase of $\alpha$, which becomes much more obvious when $\kappa$ increases. This interesting behavior is similar to that seen for the ($3+1$)-dimensional Gauss-Bonnet superconductors with the backreactions, where the critical temperature first decreases then increases as the curvature correction tends towards the Chern-Simons value in a backreaction dependent fashion \cite{BarclayGregory}. However, this trend will become less obvious if we set $m^2L_{\rm eff}^2=0$, i.e., the critical temperature $T_c$ always decreases as $\alpha$ increases for all cases considered here (we even checked the numerical data for $\kappa=0.50$), just as shown in the right-down panel. Obviously, the combination of the Gauss-Bonnet gravity and the backreaction provides richer physics in the phase transition.

\begin{figure}[ht]
\includegraphics[scale=0.57]{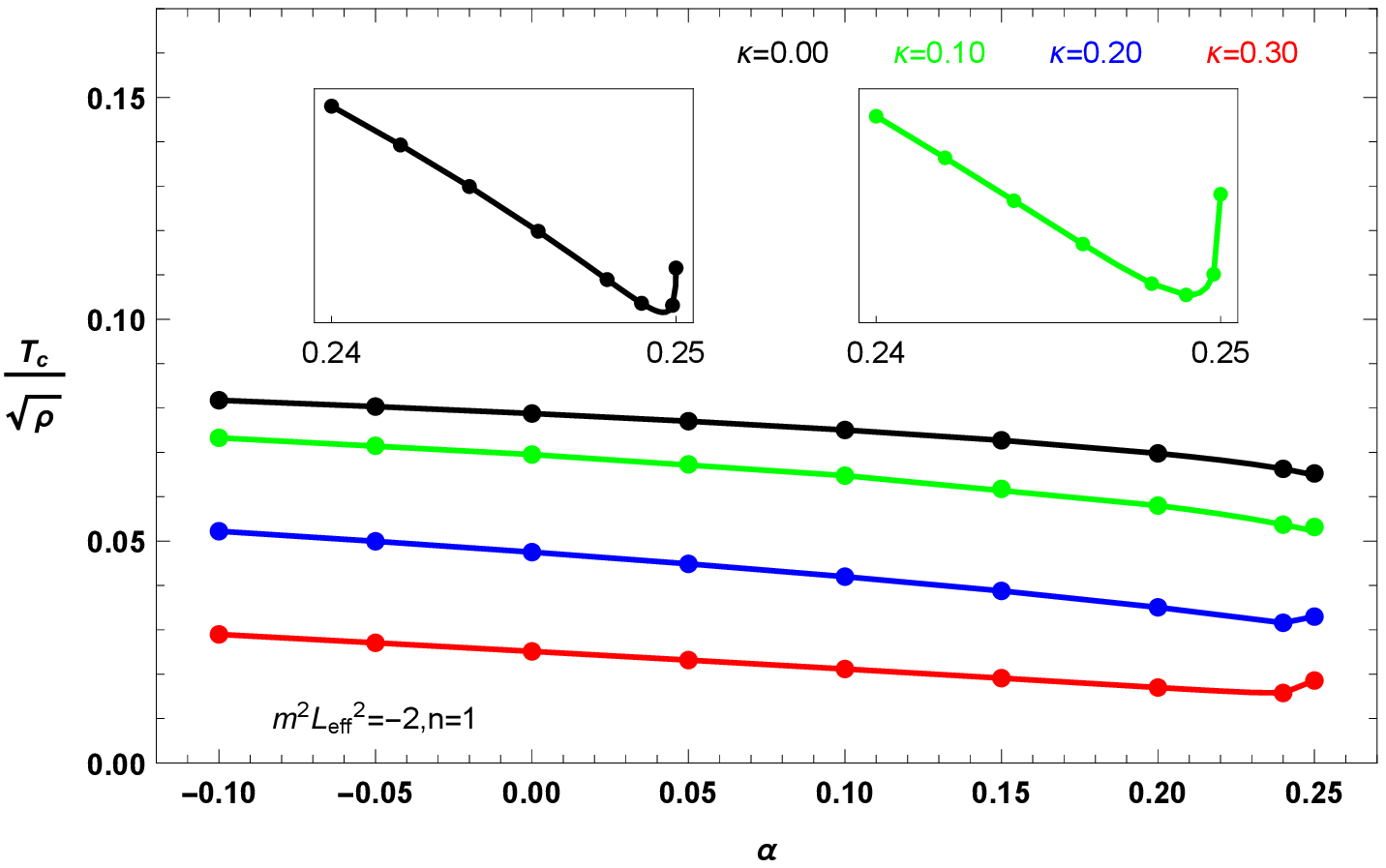}
\includegraphics[scale=0.57]{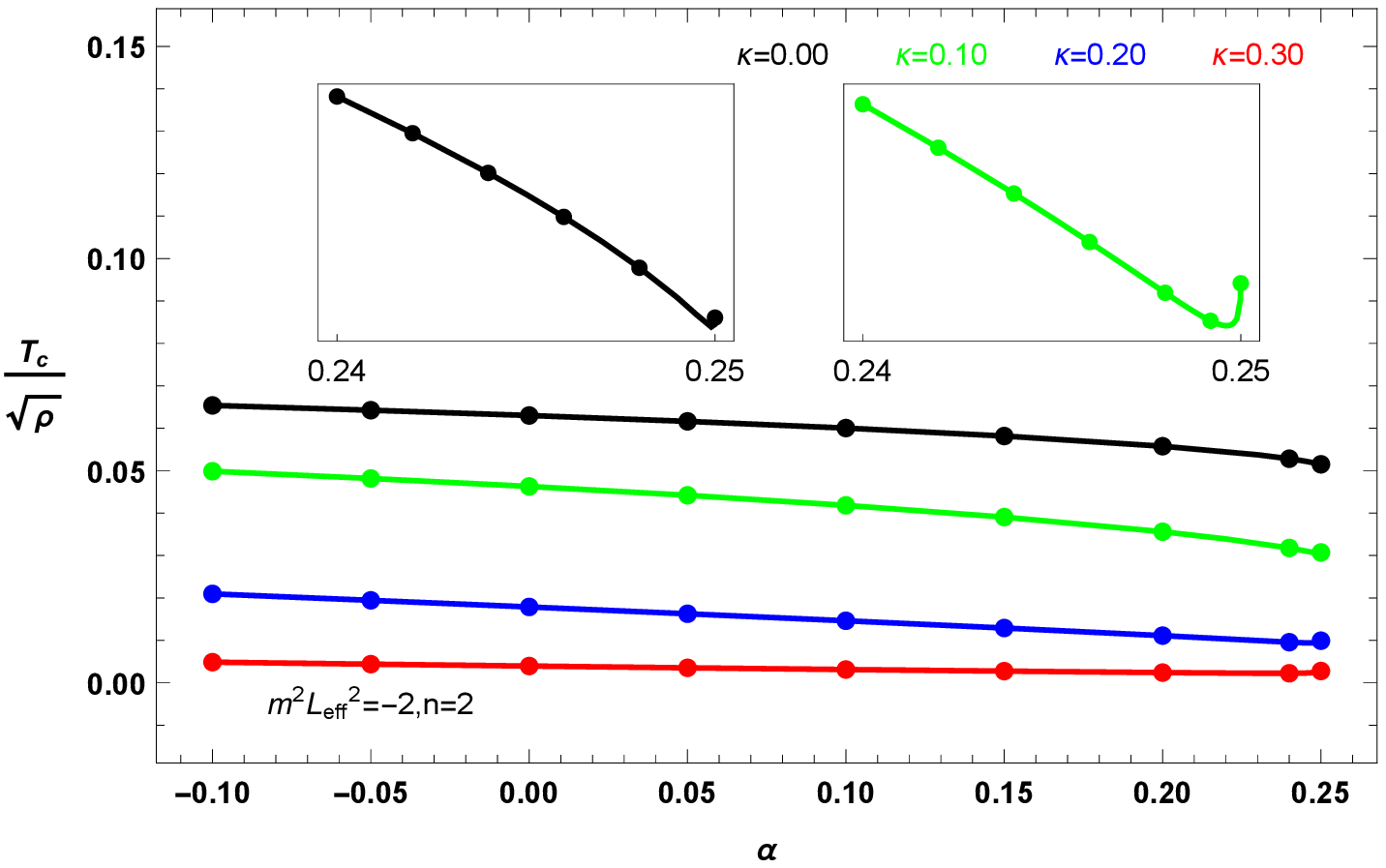}
\caption{\label{figure3}
(color online) The critical temperature $T_{c}$ of the scalar operator ${\cal O}_{+}$ as a function of the Gauss-Bonnet parameter for the fixed mass $m^2L_{\rm eff}^2=-2$ with different backreaction parameters $\kappa$ in the first ($n=1$, left) and second ($n=2$, right) states. In each panel, the four lines from top to bottom correspond to the increasing backreaction parameters, i.e., $\kappa=0.00$ (black), $0.10$ (green), $0.20$ (blue) and $0.30$ (red), respectively.}
\end{figure}

For the ground state, we have obtained the influence of the curvature correction on the critical temperature in the $(2+1)$-dimensional holographic superconductors with the backreactions for the scalar operator ${\cal O}_{+}$. We can expect
this tendency to be the same even in excited states. Thus, fixing the mass of the scalar field by $m^2L_{\rm eff}^2=-2$, in Fig. \ref{figure3} we give the critical temperature $T_c$ of the scalar operator ${\cal O}_{+}$ as a function of the Gauss-Bonnet parameter with different backreaction parameters for the first ($n=1$) and second ($n=2$) states, which clearly shows that the excited state has a lower critical temperature than the corresponding ground state. Interestingly, for the first two excited states, the critical temperature $T_{c}$ decreases first and then increases as $\alpha$ tends towards the Chern-Simons value in a backreaction dependent fashion, similarly to the case of the ground state ($n=0$) in the left-up panel of Fig. \ref{figure2}. However, this upwarping phenomenon near the Chern-Simons limit becomes less obvious when $n$ increases, for example in the case of $n=2$ shown here, the critical temperature $T_c$ almost decreases as $\alpha$ increases for $\kappa=0.00$. So we conclude that, just as the decrease of $\kappa$ or increase of $m^2L_{eff}^2$, the increase of the number of nodes $n$ will weaken the subtle effect of the curvature correction on the critical temperature.

\subsection{Operator ${\cal O}_{-}$}

Now we move to study the scalar operator ${\cal O}_{-}$ by imposing the boundary condition $\psi_{+}=0$. For the range $-9/4<m^{2}L_{\rm eff}^2<-5/4$ where both modes of the asymptotic values of the scalar fields are normalizable \cite{HorowitzPRD78}, in this section we will set $m^2L_{\rm eff}^2=-2$ for concreteness since the other choices will not qualitatively modify our results.

\begin{figure}[ht]
\includegraphics[scale=0.65]{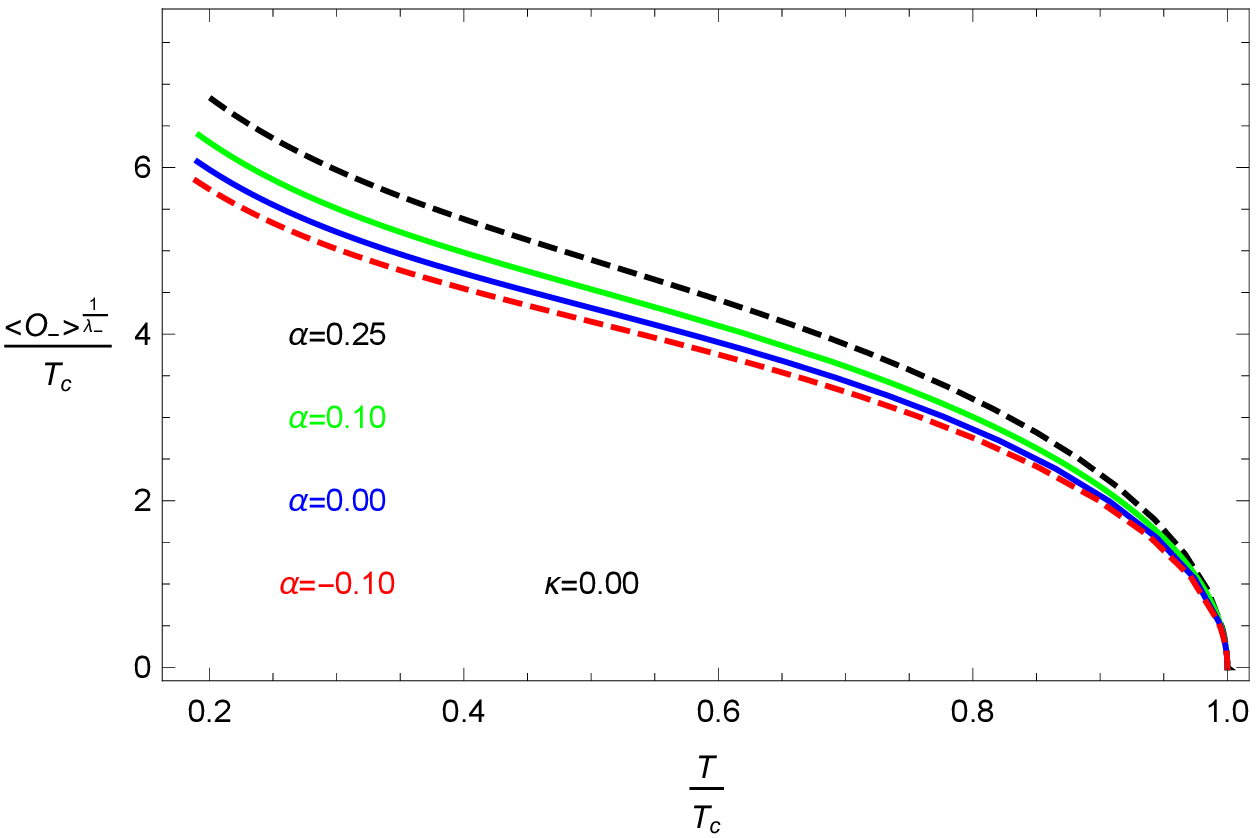}
\includegraphics[scale=0.65]{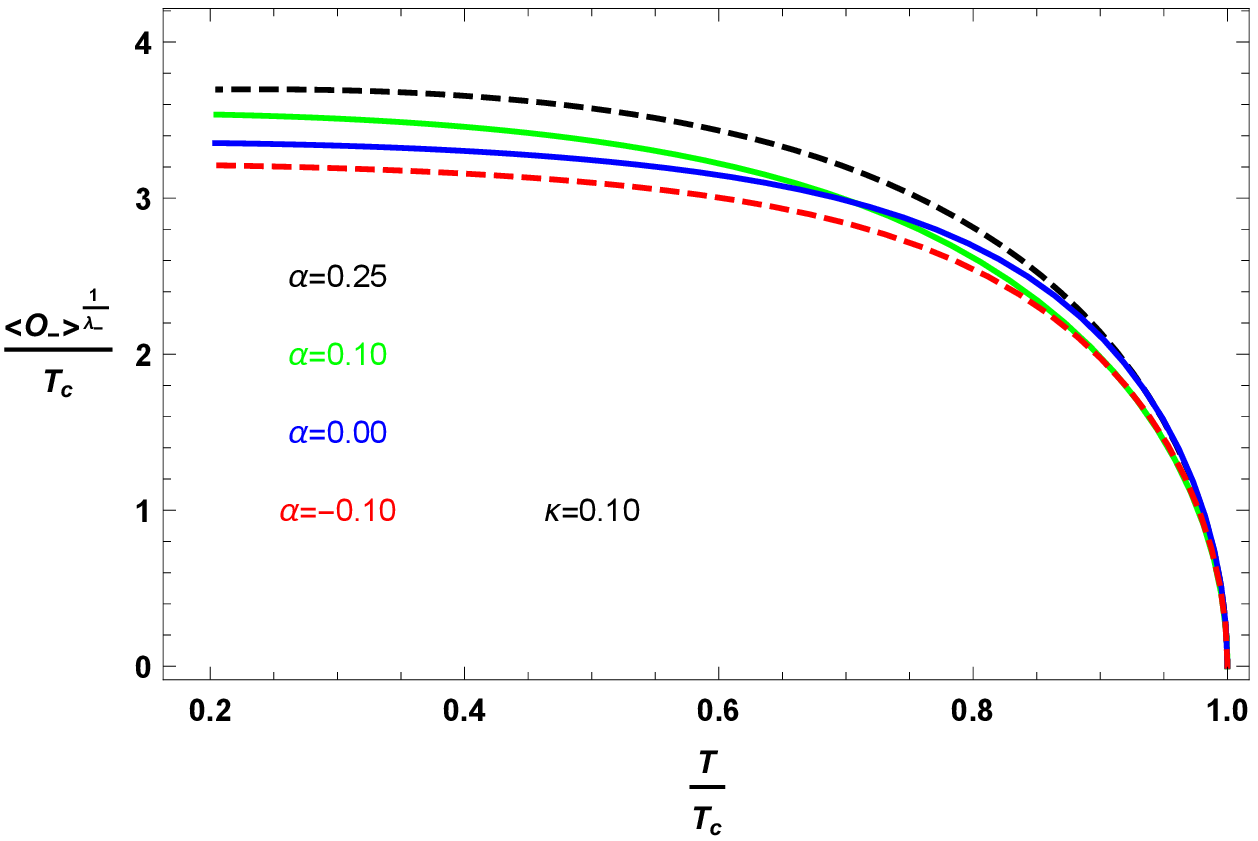}\\
\includegraphics[scale=0.65]{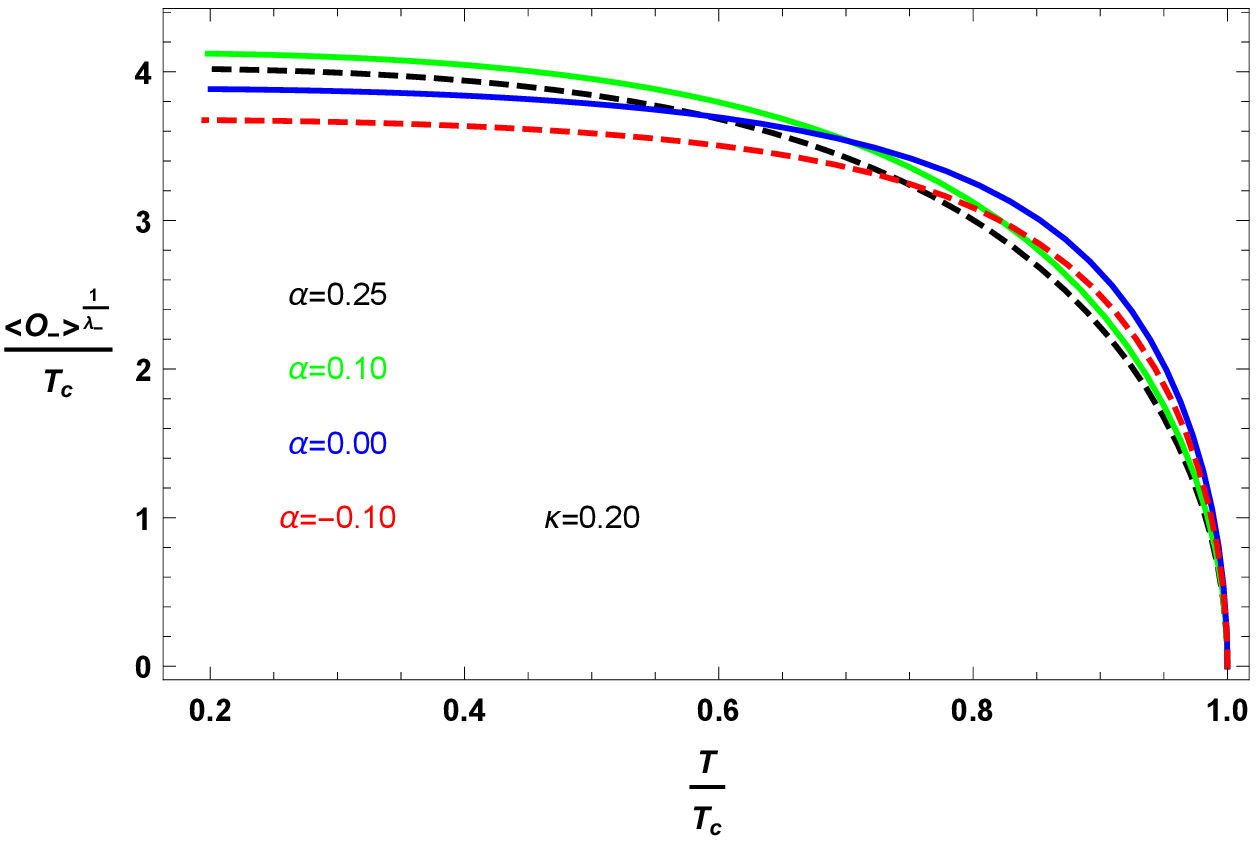}
\includegraphics[scale=0.65]{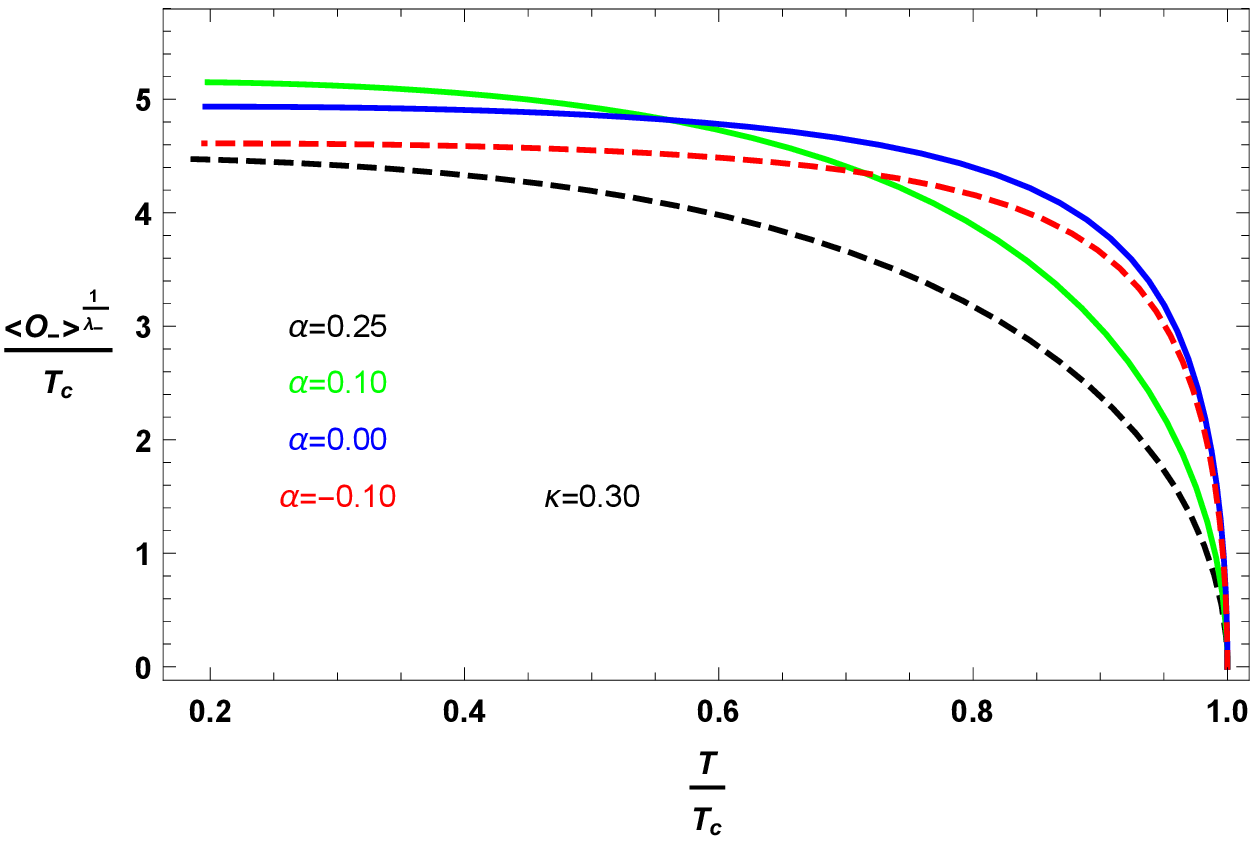}\\
\caption{\label{figure4}(color online) The condensates of the scalar operator ${\cal O}_{-}$ as a function of temperature for the fixed mass of the scalar field $m^{2}L_{\rm eff}^{2}=-2$ with different Gauss-Bonnet parameters $\alpha$ and backreaction parameters $\kappa$ in the ground state. In each panel, the four lines correspond to the increasing Gauss-Bonnet parameter, i.e., $\alpha=-0.10$ (red and dashed), $0.00$ (blue), $0.10$ (green) and $0.25$ (black and dashed), respectively.}
\end{figure}

In Fig. \ref{figure4}, we plot the condensates of the scalar operator ${\cal O}_{-}$ as a function of temperature with various Gauss-Bonnet parameters and backreaction parameters in the ground state. For the left-up panel, similar to those for the standard $(2+1)$-dimensional holographic superconductor model in the probe limit \cite{HartnollPRL101}, the curves for the operator ${\cal O}_{-}$ with $\kappa=0.00$ will diverge at low temperature. But taking the backreactions of the spacetime into account, i.e., $\kappa=0.10$, $0.20$ and $0.30$, the curve goes to a constant at zero temperature for each $\alpha$, which is similar to the BCS theory and being used to describe the superconductivity. By fitting these curves near the critical point, we have $\langle {\cal O}_{-}\rangle\sim (1-T/T_{c})^{1/2}$. This behavior is reminiscent of that seen for the scalar operator ${\cal O}_{+}$, so we conclude that the phase transition of the $4D$ Gauss-Bonnet holographic superconductors with the backreactions is typical of second order one with the mean field critical exponent $1/2$.

\begin{figure}[ht]
\includegraphics[scale=0.65]{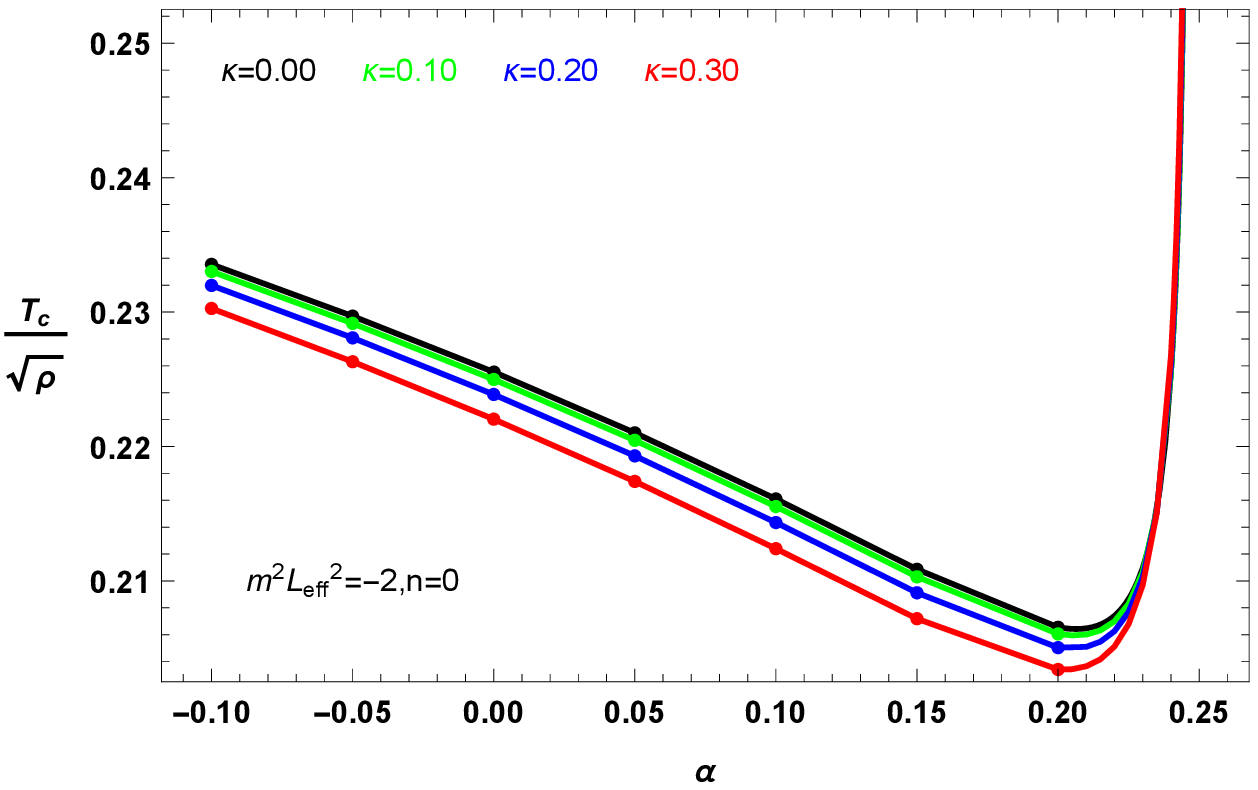}
\includegraphics[scale=0.65]{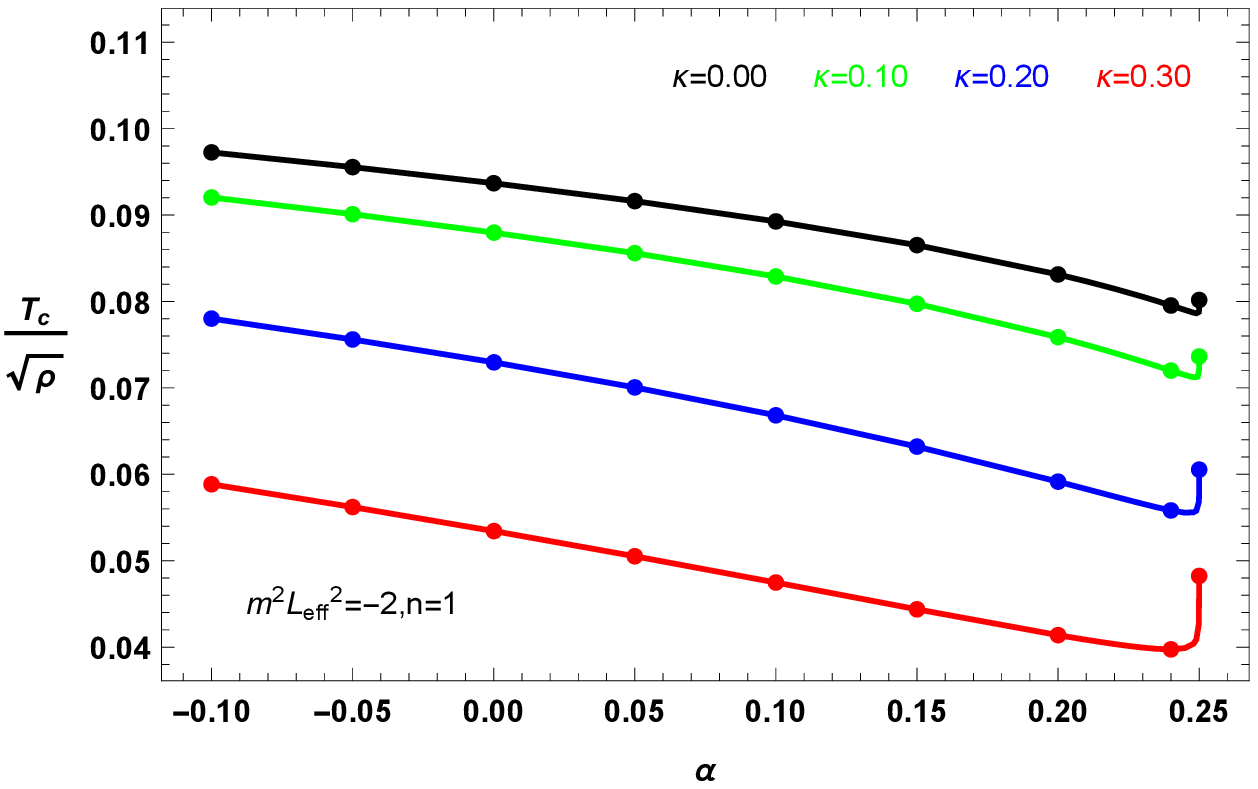}\\
\includegraphics[scale=0.57]{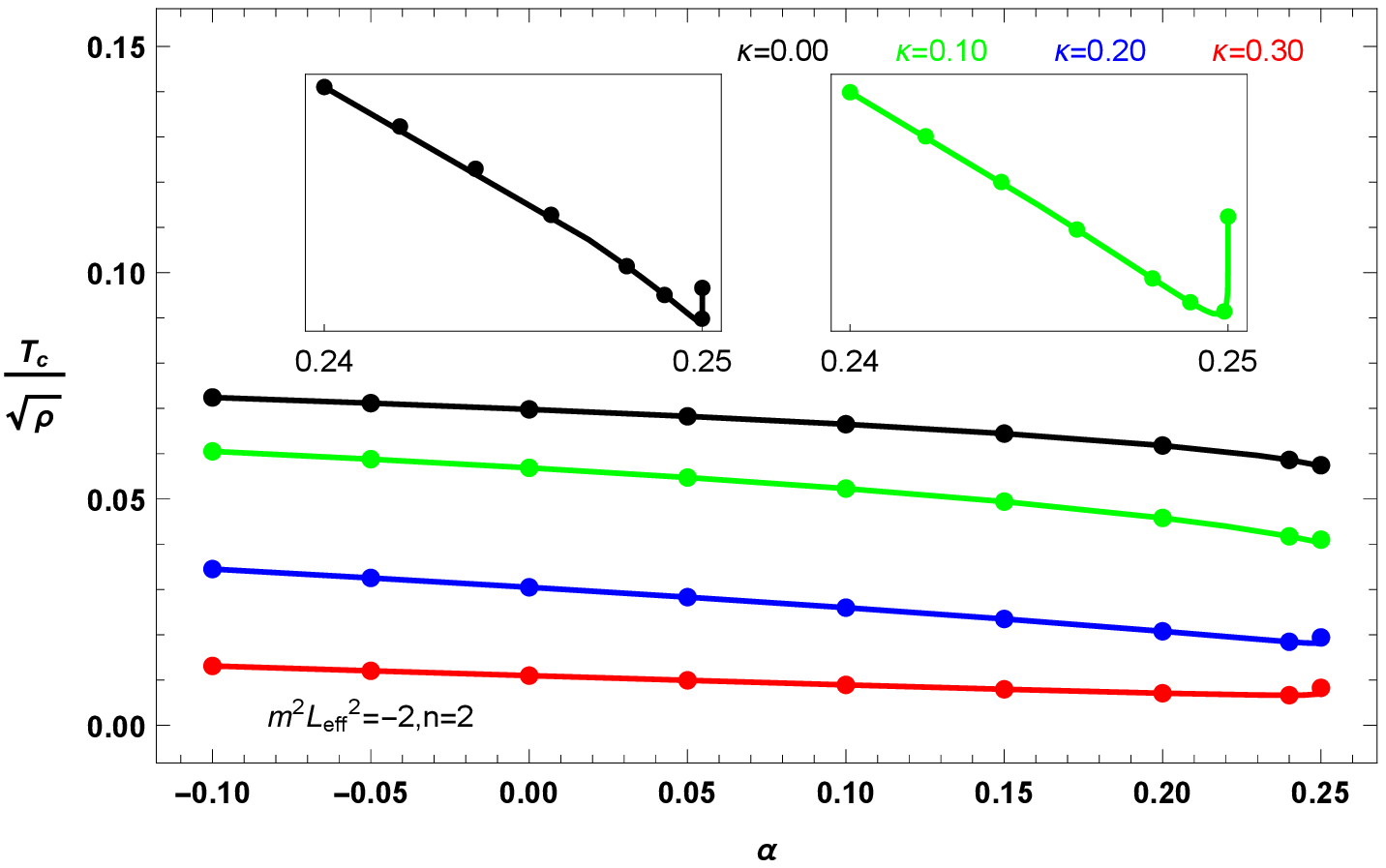}
\includegraphics[scale=0.57]{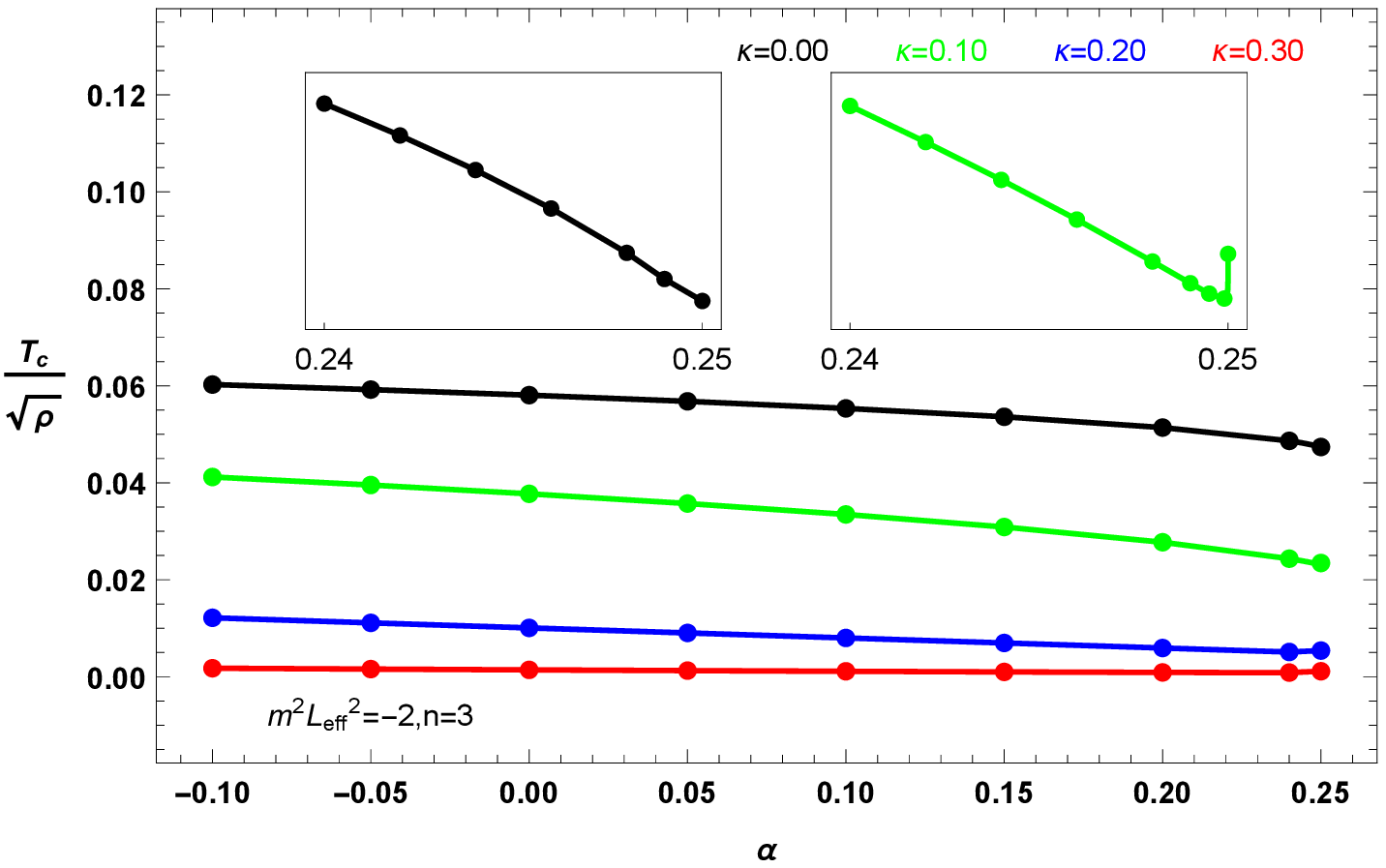}
\caption{\label{figure5} (color online) The critical temperature $T_{c}$ of the scalar operator ${\cal O}_{-}$ as a function of the Gauss-Bonnet parameter for the fixed mass $m^2L_{\rm eff}^2=-2$ with different backreaction parameters $\kappa$ in the ground ($n=0$), first ($n=1$), second ($n=2$) and third ($n=3$) states. In each panel, the four lines from top to bottom correspond to the increasing backreaction parameters, i.e., $\kappa=0.00$ (black), $0.10$ (green), $0.20$ (blue) and $0.30$ (red), respectively.}
\end{figure}

In Fig. \ref{figure5}, we present the critical temperature $T_c$ of the scalar operator ${\cal O}_{-}$ as a function of the Gauss-Bonnet parameter with different backreaction parameters for the first four lowest-lying modes, i.e., the ground ($n=0$), first ($n=1$), second ($n=2$) and third ($n=3$) states, which tells us that there exists a lower critical temperature in the corresponding excited state. Similar to the scalar operator ${\cal O}_{+}$, we observe the effect of $\alpha$, which initially lowers the critical temperature $T_{c}$, then increases it again for $\alpha$ approaching the Chern-Simons value. It is also interesting to note that, from Figs. \ref{figure3} and \ref{figure5}, this upwarping phenomenon near the Chern-Simons limit becomes less obvious with the increase of $n$, for example the case of $n=3$, which shows that increasing $n$ really weakens the subtle effect of the Gauss-Bonnet coupling $\alpha$ on the critical temperature $T_{c}$.

\section{Conductivity}

Now we want to know the influence of the curvature correction on the conductivity in the $(2+1)$-dimensional holographic superconductors with the backreactions. Assuming the time-dependent perturbation with zero momentum $\delta A_{x}=A_{x}(r)e^{-i\omega t}dx$ and $g_{tx}=g(r)e^{-i\omega t}$ \cite{HartnollJHEP12}, we have
\begin{eqnarray}
A_{x}^{\prime\prime}+\left(\frac{f^\prime}{f}-\frac{\chi^\prime}{2}\right)A_{x}^\prime
+\left(\frac{\omega^2}{f^2}e^{\chi}-\frac{2\psi^{2}}{f}\right)A_{x}
+\frac{\phi^\prime}{f}e^{\chi}\left(g^\prime-\frac{2g}{r}\right)=0,
\label{MaxwellEquation}
\end{eqnarray}
\begin{eqnarray}
g^\prime-\frac{2g}{r}+\frac{2\kappa^{2}r^{2}\phi^\prime}{r^2-\alpha f}A_{x}=0, \label{MetricPerturbation}
\end{eqnarray}
which leads to the equation of motion for the perturbed Maxwell field
\begin{eqnarray}
A_{x}^{\prime\prime}+\left(\frac{f'}{f}-\frac{\chi'}{2}\right)A_{x}^\prime
+\left\{\left[\frac{\omega^2}{f^2}-\frac{2\kappa^2r^2\phi'^2}{(r^2-\alpha f)f}\right]e^\chi-\frac{2\psi^2}{f}\right\}A_x=0.
\end{eqnarray}
Near the horizon, the ingoing wave boundary condition reads
\begin{eqnarray}
A_x(r)\sim f(r)^{-i\frac{\omega}{4\pi T_{H}}},
\end{eqnarray}
where $T_{H}$ is the Hawking temperature given by Eq. (\ref{HawkingT}), and in the asymptotic AdS region
\begin{eqnarray}
A_{x}(r)=A^{(0)}_{x}(r)+\frac{A^{(1)}_{x}(r)}{r},~~g(r)=r^{2}g^{(0)}+\frac{g^{(1)}}{r}.
\end{eqnarray}
Using the AdS/CFT dictionary, we can obtain the conductivity of the dual superconductor \cite{HartnollPRL101,HartnollJHEP12}
\begin{eqnarray}
\sigma=-\frac{iA_x^{(1)}(r)}{\omega A_x^{(0)}(r)}.
\end{eqnarray}
For different values of the Gauss-Bonnet parameter $\alpha$ and backreaction parameter $\kappa$, we concentrate on the scalar operator ${\cal O}_{+}$ and obtain the conductivity by solving the Maxwell equation numerically.

\begin{figure}[ht]
\includegraphics[scale=0.43]{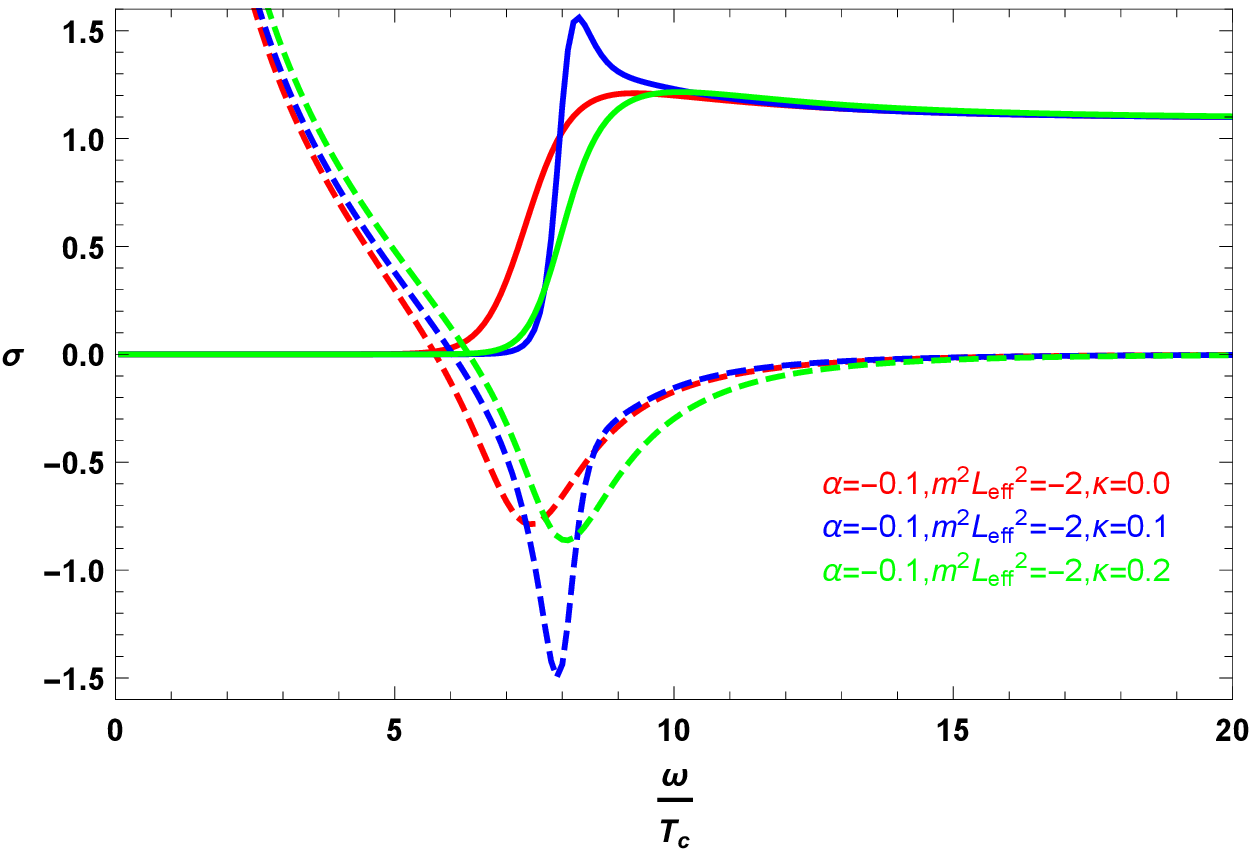}
\includegraphics[scale=0.43]{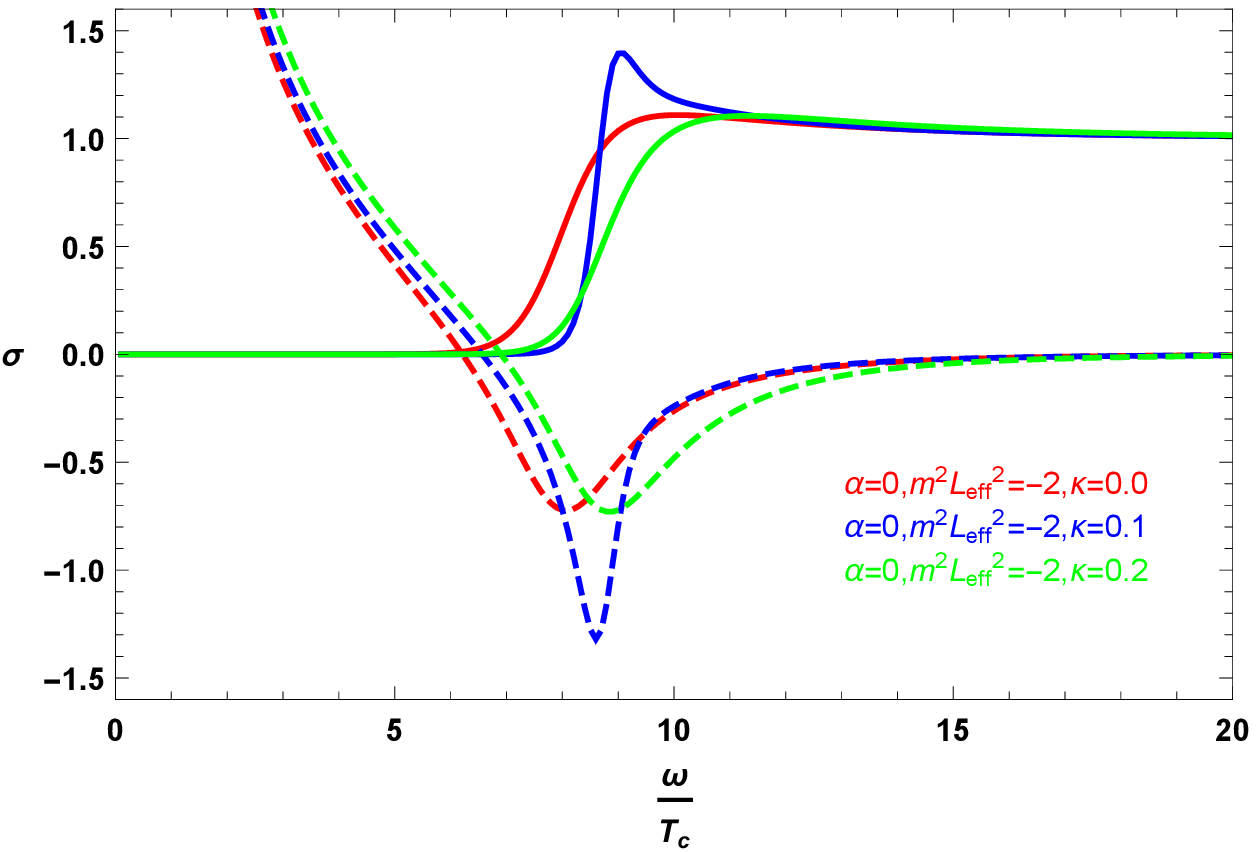}
\includegraphics[scale=0.43]{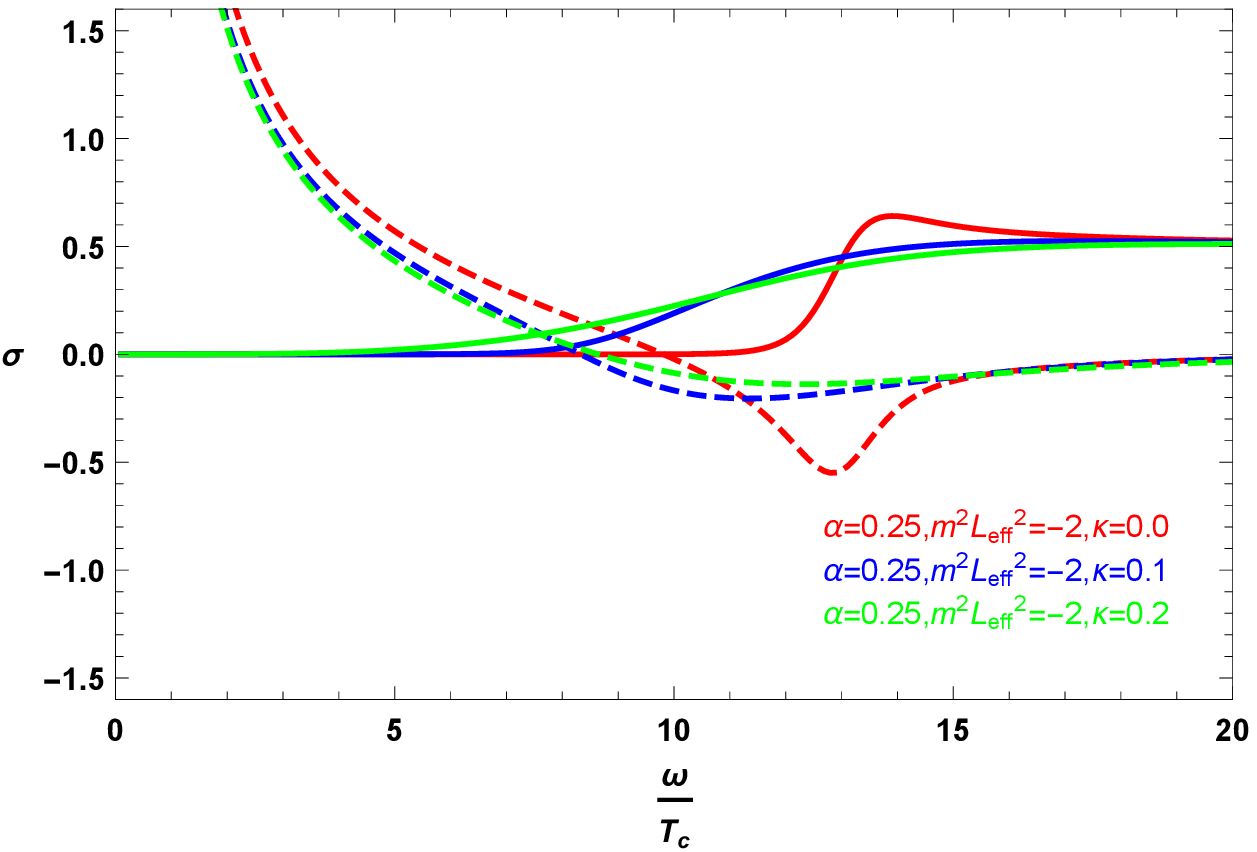}\\
\includegraphics[scale=0.43]{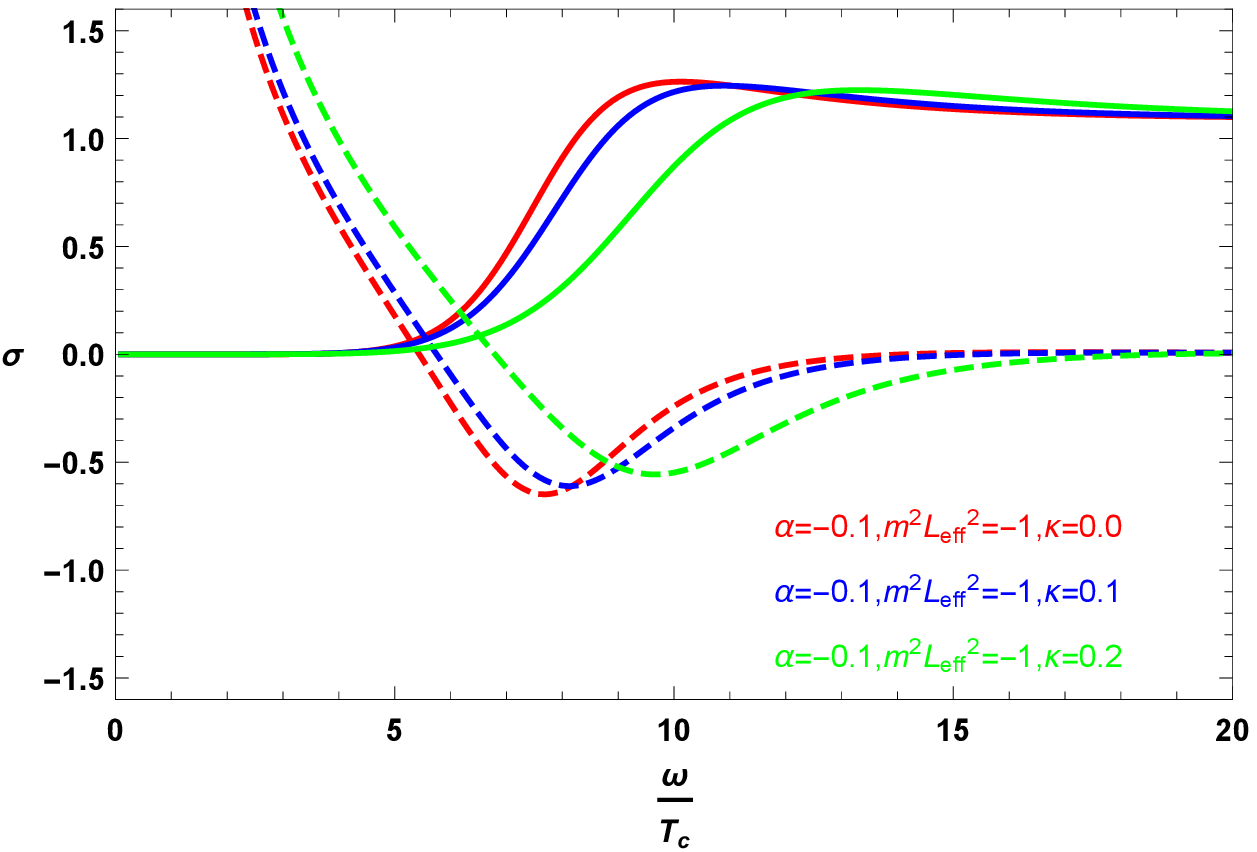}
\includegraphics[scale=0.43]{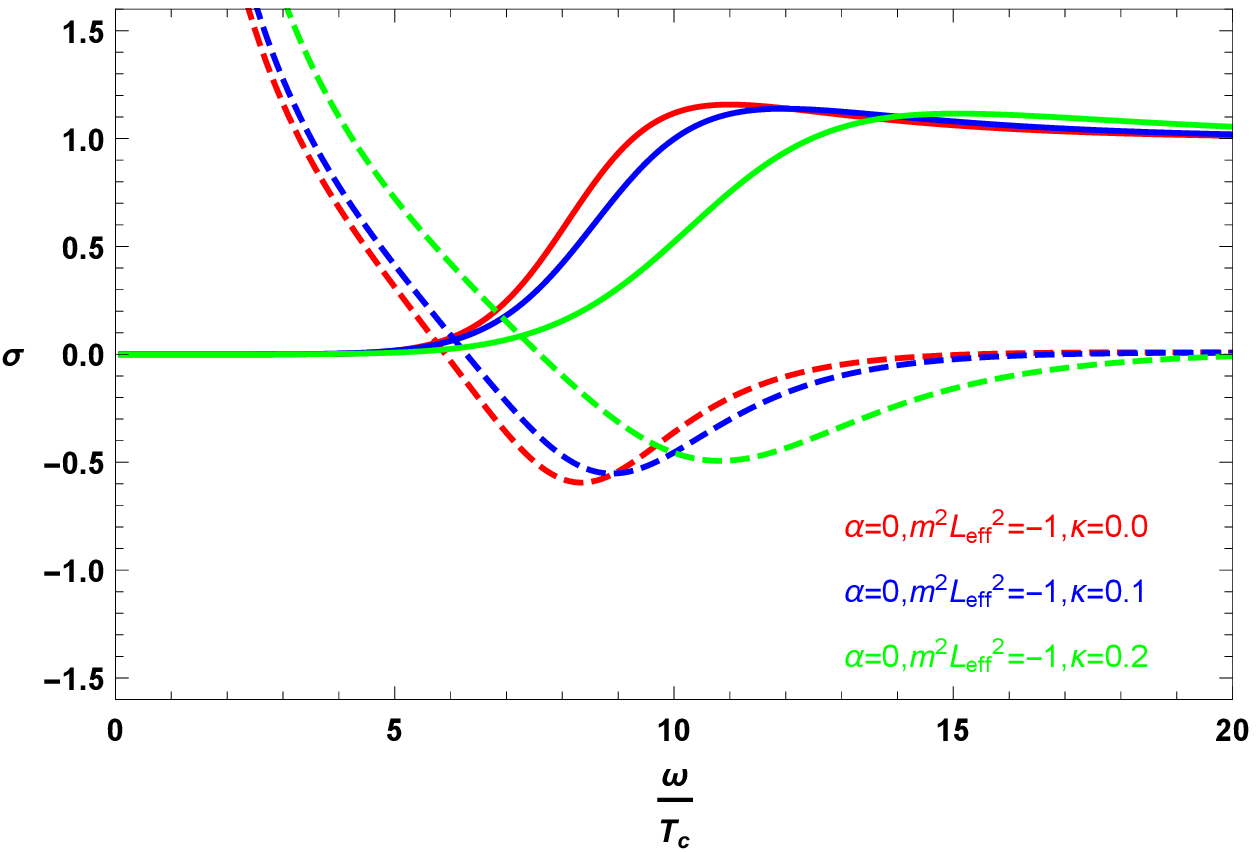}
\includegraphics[scale=0.43]{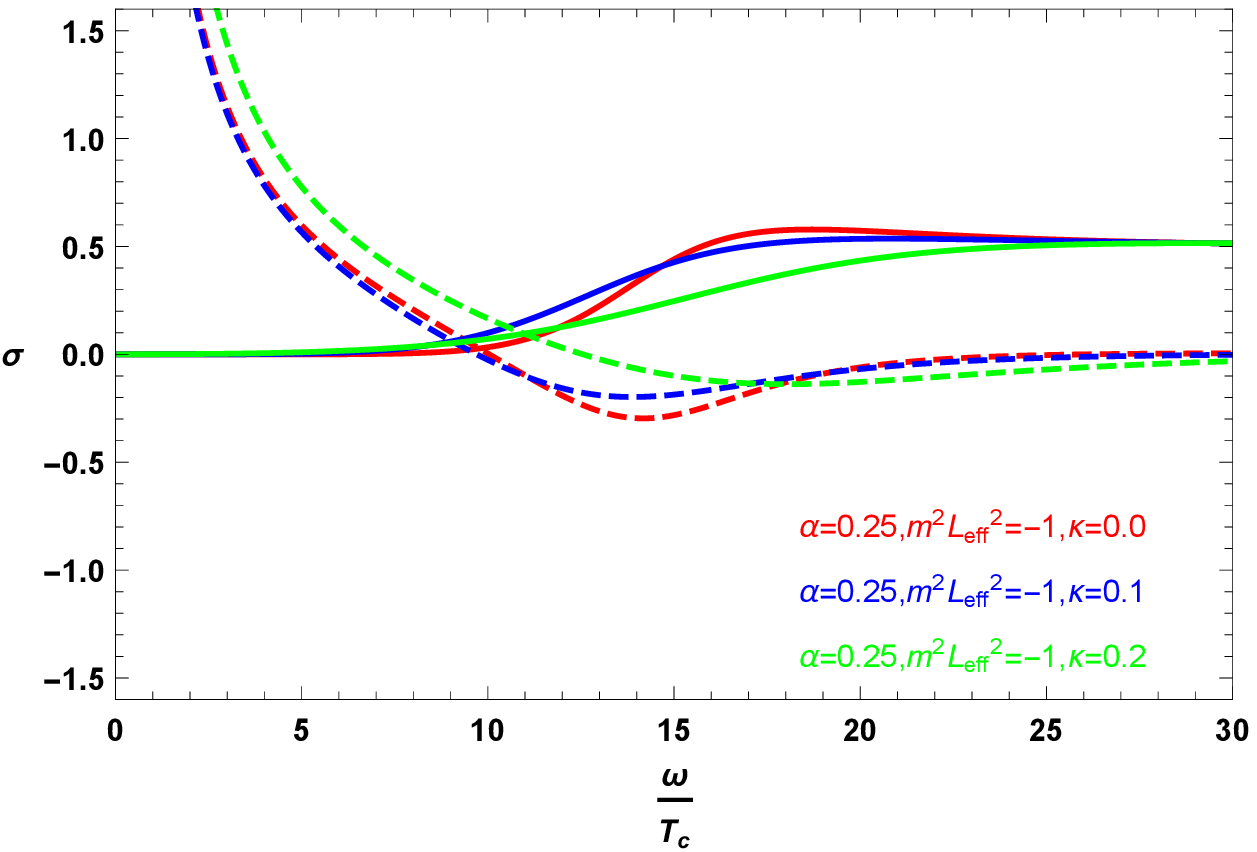}\\
\includegraphics[scale=0.43]{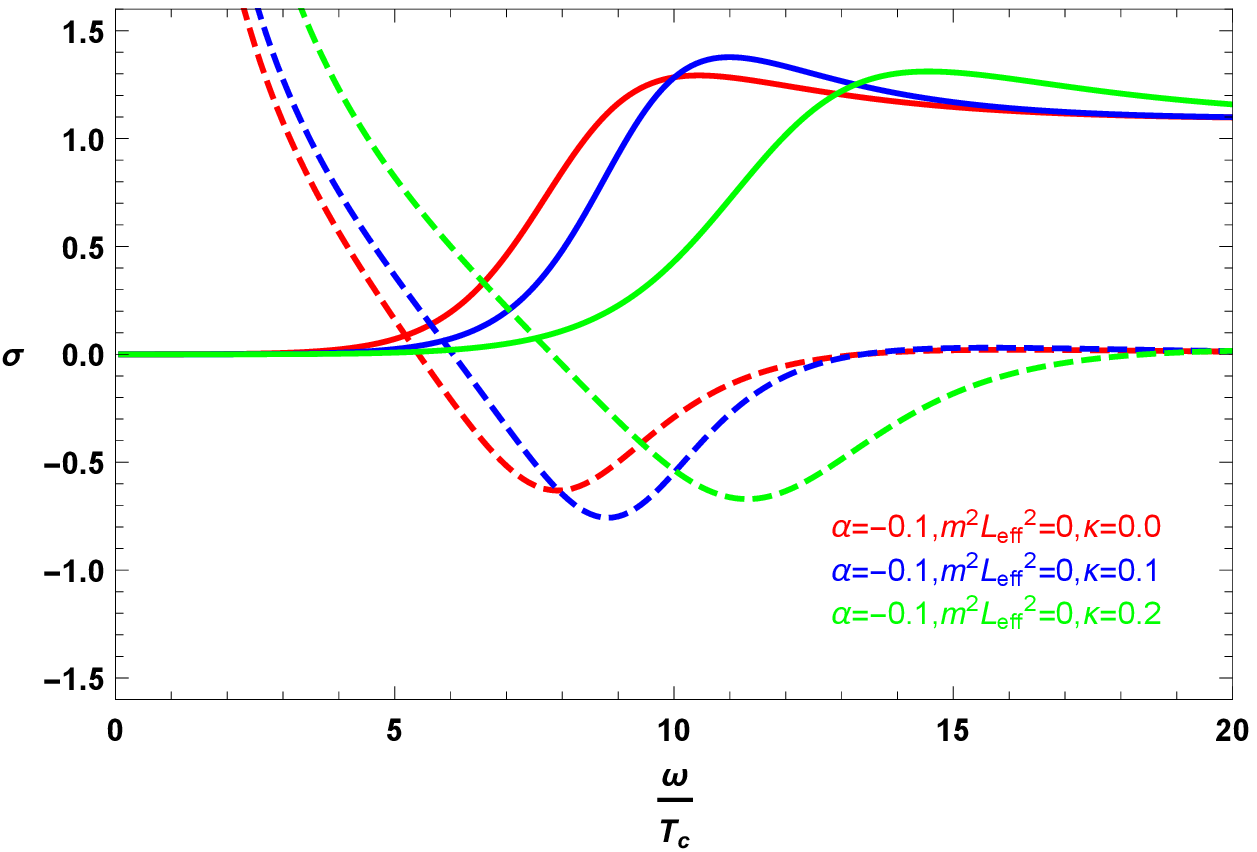}
\includegraphics[scale=0.43]{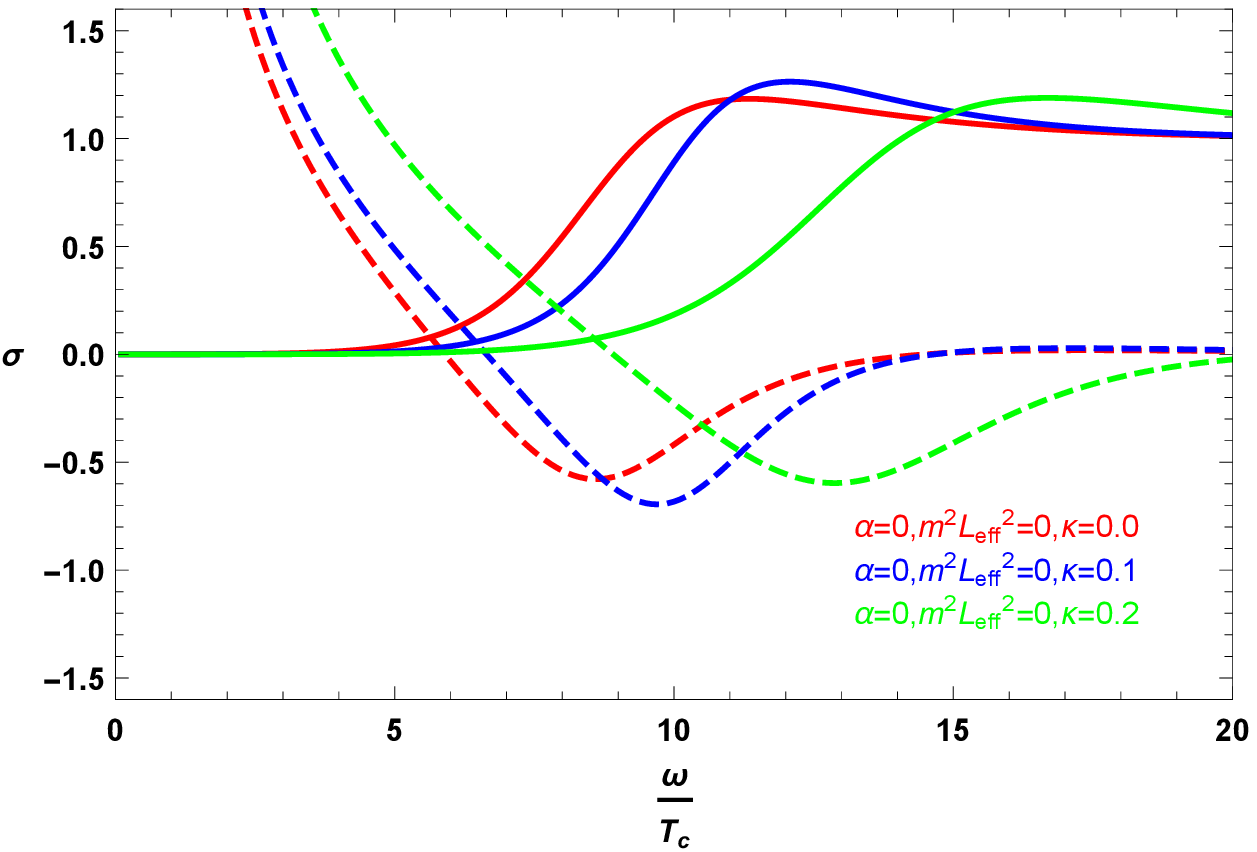}
\includegraphics[scale=0.43]{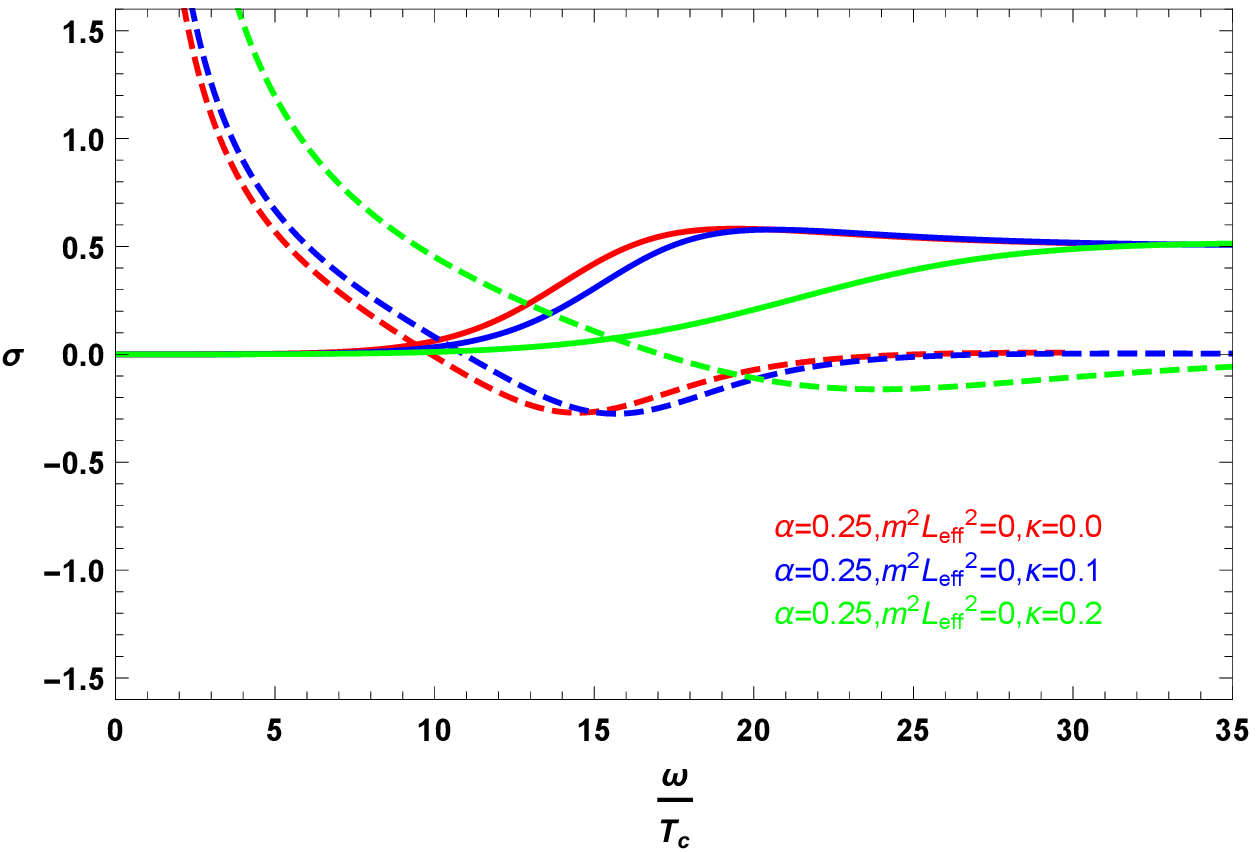}\\
\caption{\label{Conductivityn0} (color online) Conductivity of the ($2+1$)-dimensional superconductors for the fixed masses of the scalar field $m^{2}L_{\rm eff}^{2}=-2$, $-1$ and $0$ with different Gauss-Bonnet parameters $\alpha$ and backreaction parameters $\kappa$ in the ground state. In each panel, the solid line and dashed line represent the real part and imaginary part of the conductivity $\sigma(\omega)$, respectively.}
\end{figure}

For the ground state in the ($3+1$)-dimensional Gauss-Bonnet superconductors with the backreactions, it is shown that increasing either Gauss-Bonnet parameter $\alpha$ or backreaction parameter $\kappa$ increases $\omega_g/T_c$ \cite{BarclayGregory}, where the gap frequency $\omega_g$ is defined as the frequency minimizing $Im[\sigma(\omega)]$ \cite{HorowitzPRD78}. Naturally, we expect this tendency to be the same even in ($2+1$)-dimensions. In Fig. \ref{Conductivityn0}, we plot the frequency dependent conductivity $\sigma(\omega)$ of the ($2+1$)-dimensional superconductors for the fixed mass of the scalar field $m^{2}L_{\rm eff}^{2}=-2$, $-1$ and $0$ with different $\alpha$ and $\kappa$ at temperatures $T/T_c\approx0.2$, i.e., $\alpha=-0.1$, $0$, $0.25$ and $\kappa=0$, $0.1$, $0.2$, where the solid line and dashed line represent the real part and imaginary part of $\sigma(\omega)$, respectively. Obviously, for the fixed $m^{2}L_{\rm eff}^{2}$ and $\kappa$, we can see clearly that the effect of the increasing Gauss-Bonnet parameter $\alpha$ is to increase $\omega_g/T_c$, which implies that the higher curvature corrections really alter the universal relation $\omega_g/T_c\approx 8$ \cite{HorowitzPRD78} for the ($2+1$)-dimensional superconductors with the backreactions. This is similar to the effect of the Gauss-Bonnet coupling for the ($3+1$)-dimensional superconductors with the backreactions. However, it is interesting to note that, for the fixed $m^{2}L_{\rm eff}^{2}$ and $\alpha$, the backreaction has a more subtle effect on $\omega_g/T_c$ for all cases considered here, i.e., increasing backreaction parameter $\kappa$ increases $\omega_g/T_c$ except for the case of $\alpha=0.25$ with $m^{2}L_{\rm eff}^{2}=-2$ and $-1$ where $\omega_g/T_c$ decreases first and then increases when $\kappa$ increases. Comparing with the ($3+1$)-dimensional Gauss-Bonnet superconductors with the backreactions \cite{BarclayGregory}, we find that, although the underlying mechanism remains mysterious, the ($2+1$)-dimensional Gauss-Bonnet superconductors exhibit a very interesting and different feature, i.e., the gap frequency $\omega_g/T_c$ decreases first and then increases when $\kappa$ increases in a scalar mass dependent fashion for $\alpha$ approaching the Chern-Simons limit.

\begin{figure}[ht]
\includegraphics[scale=0.43]{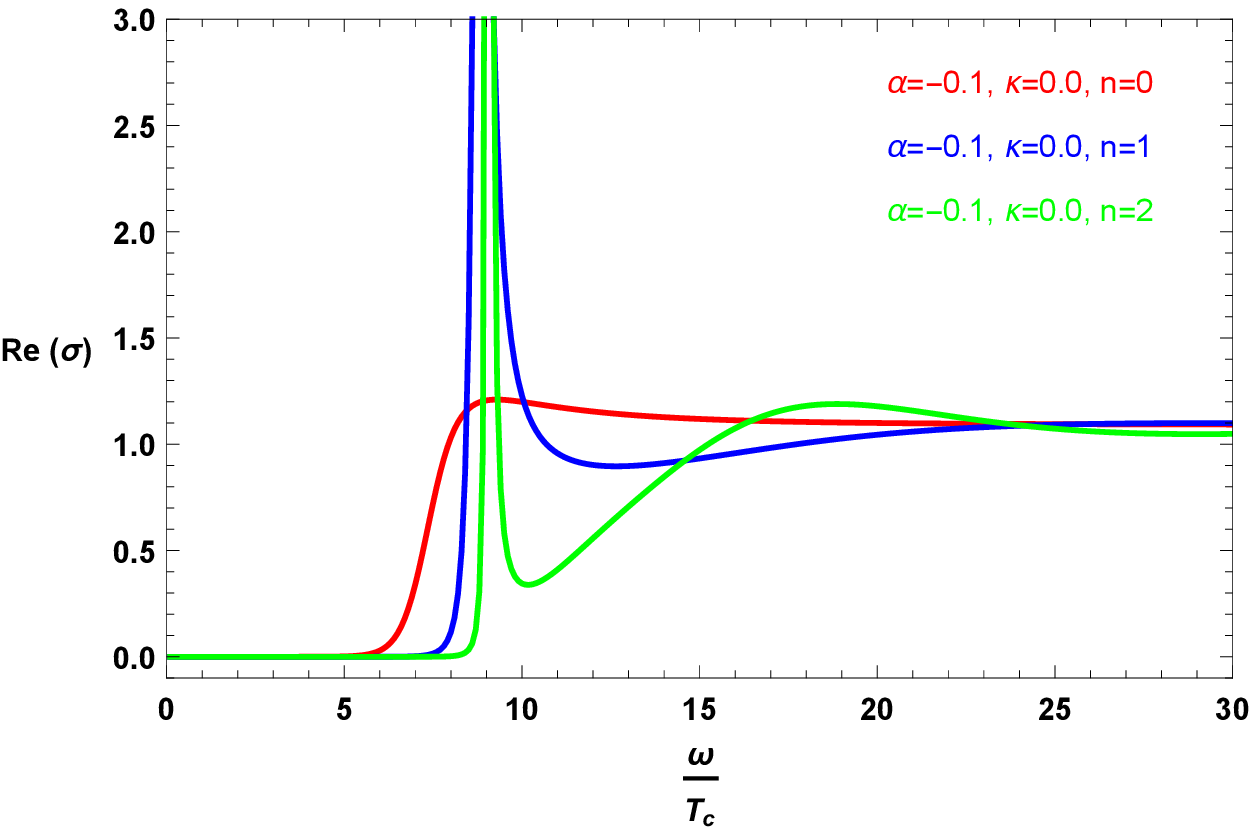}
\includegraphics[scale=0.43]{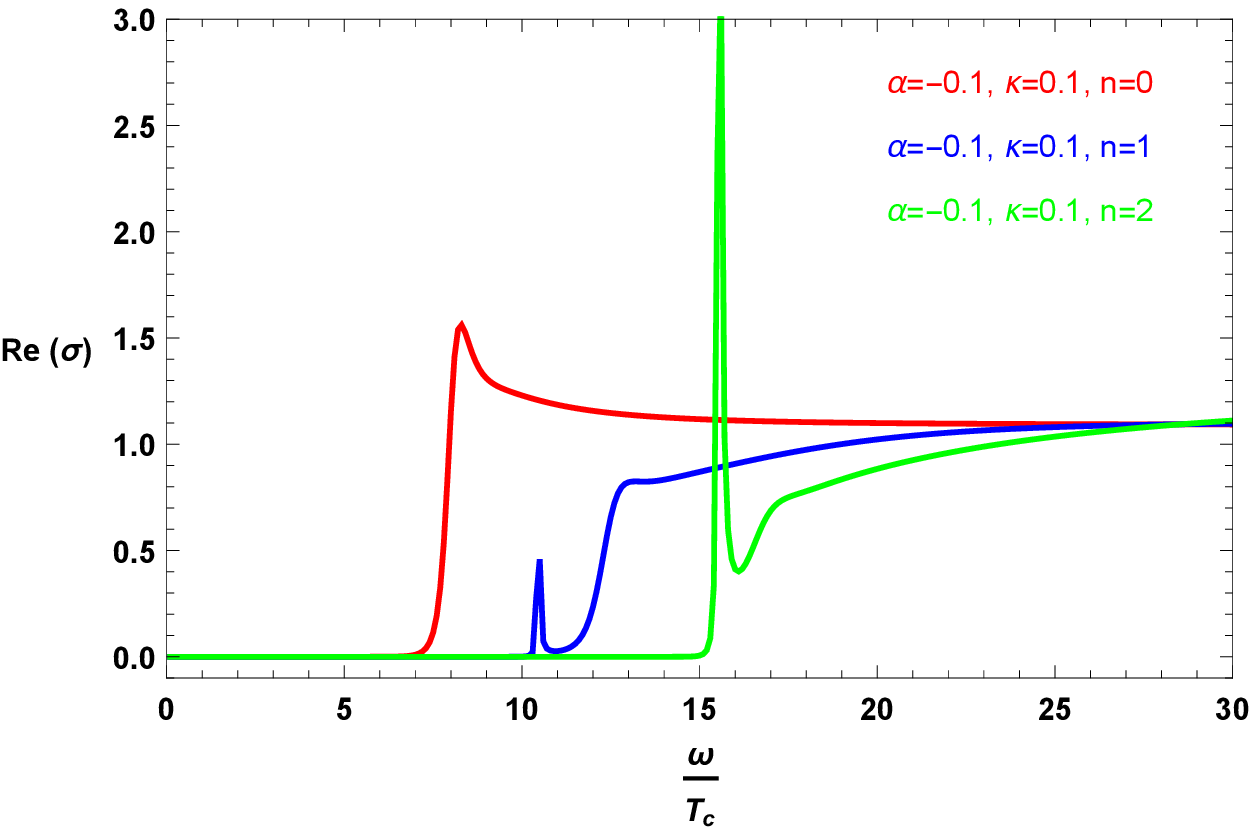}
\includegraphics[scale=0.43]{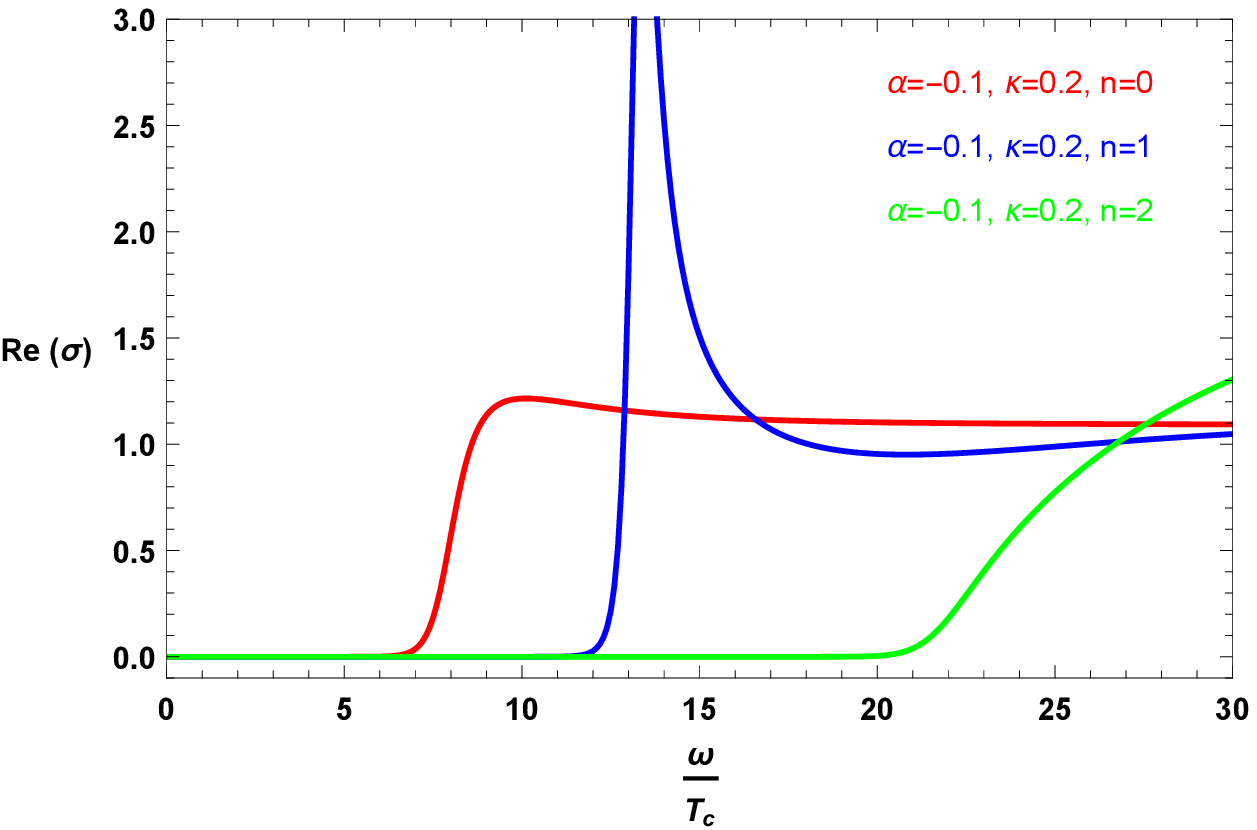}\\
\includegraphics[scale=0.43]{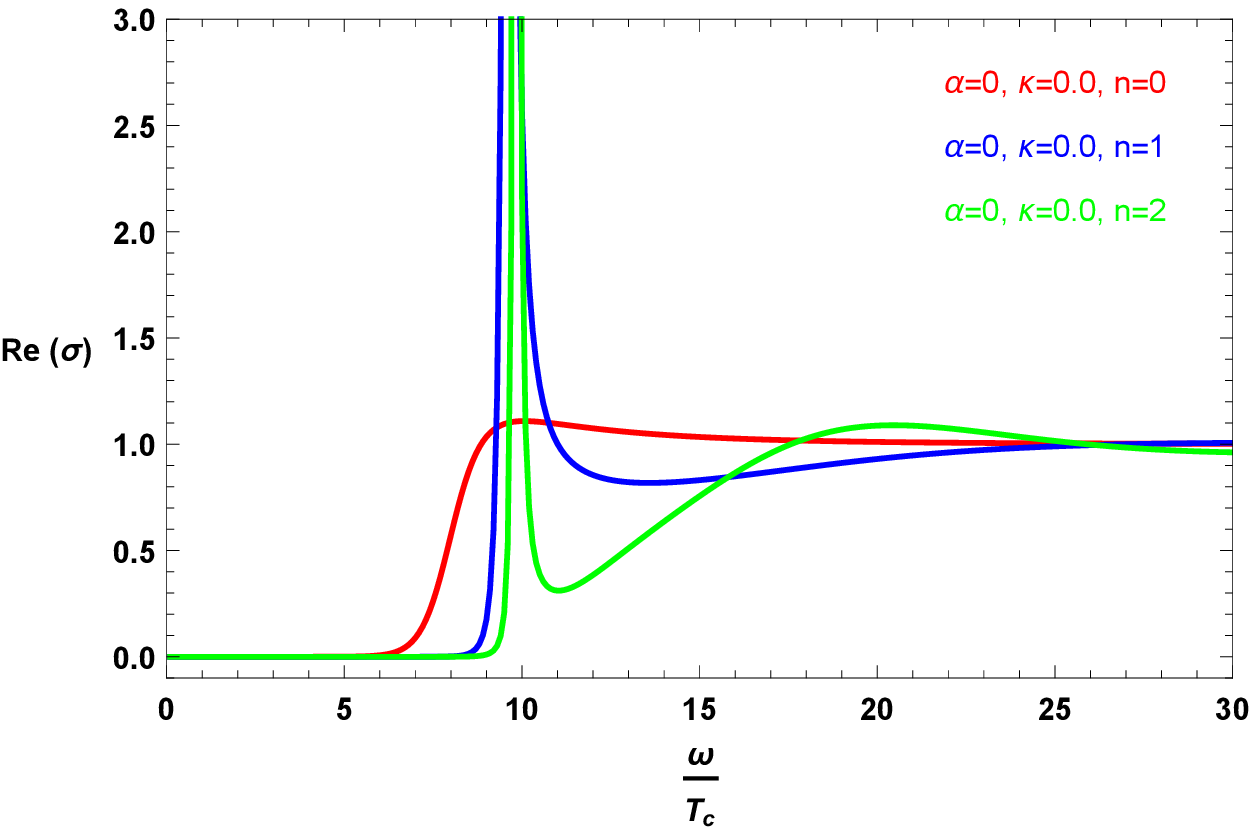}
\includegraphics[scale=0.43]{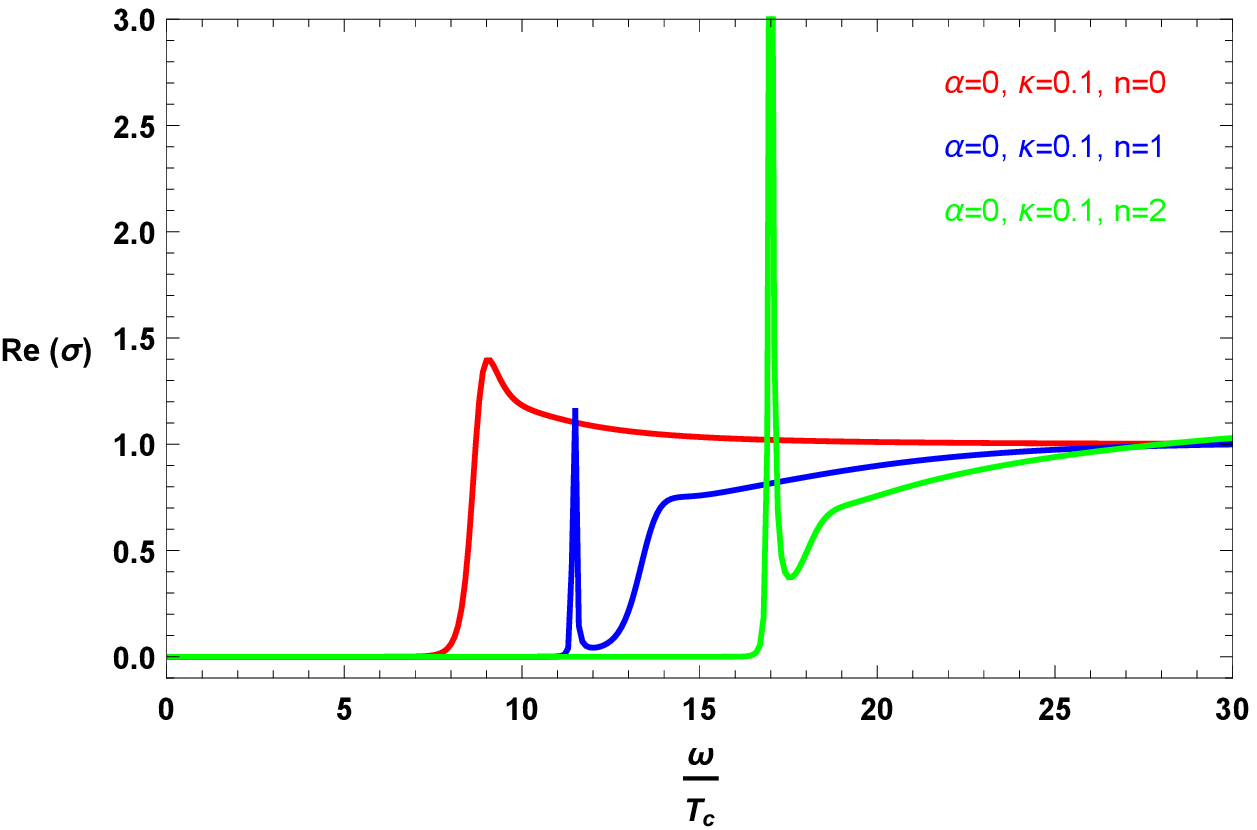}
\includegraphics[scale=0.43]{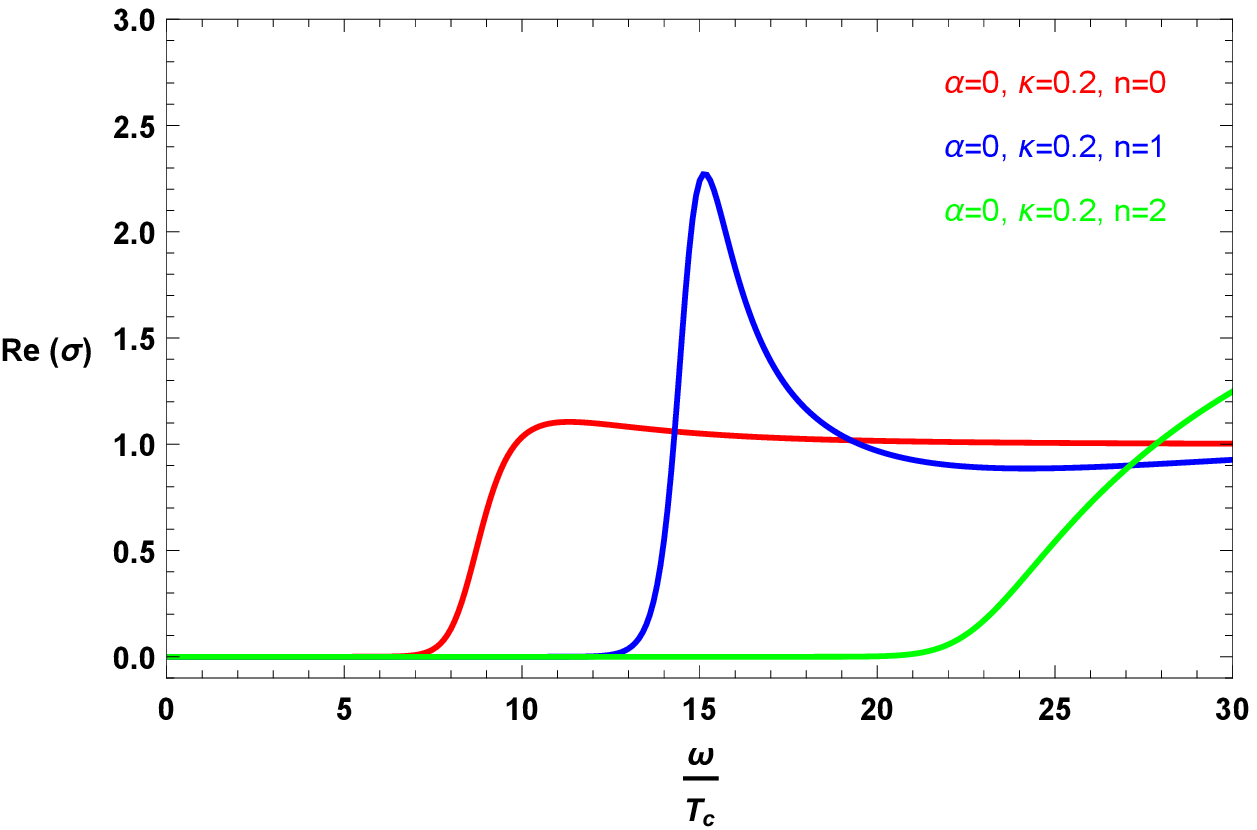}\\
\includegraphics[scale=0.43]{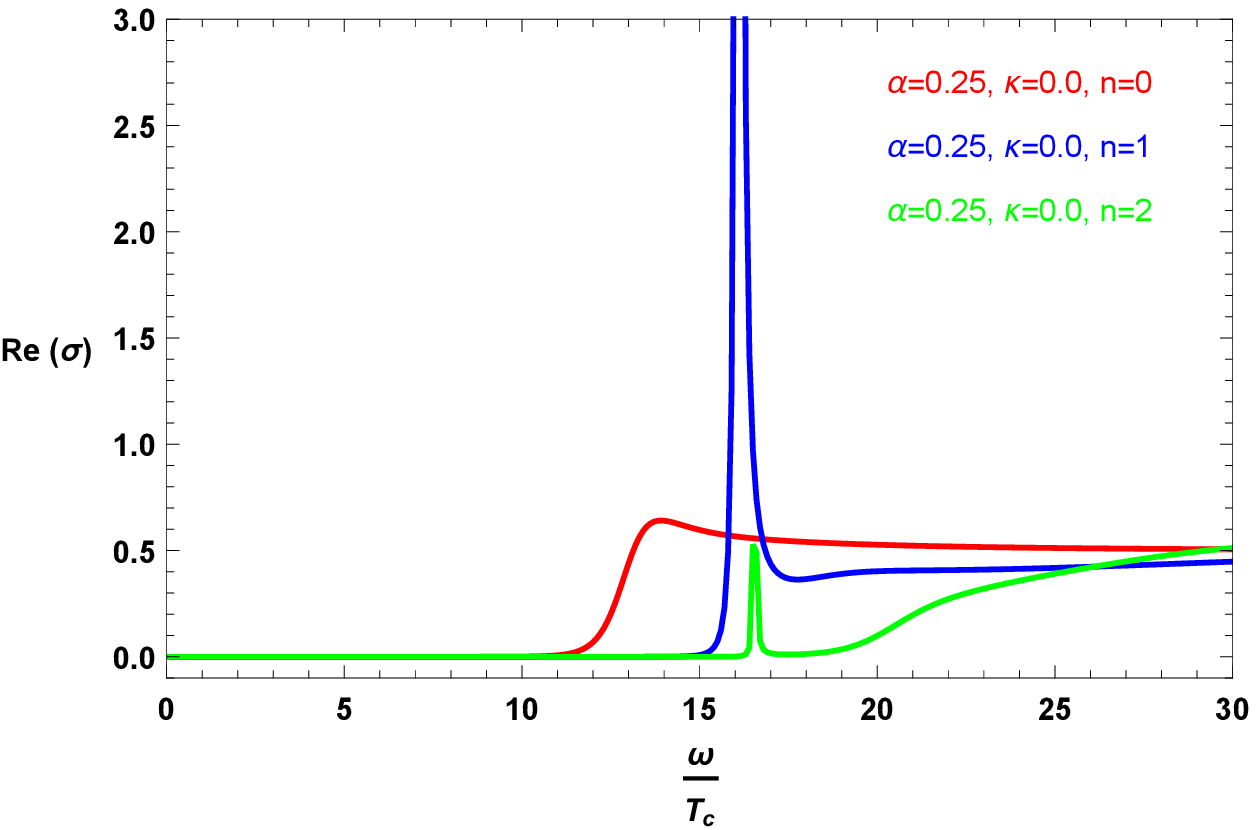}
\includegraphics[scale=0.43]{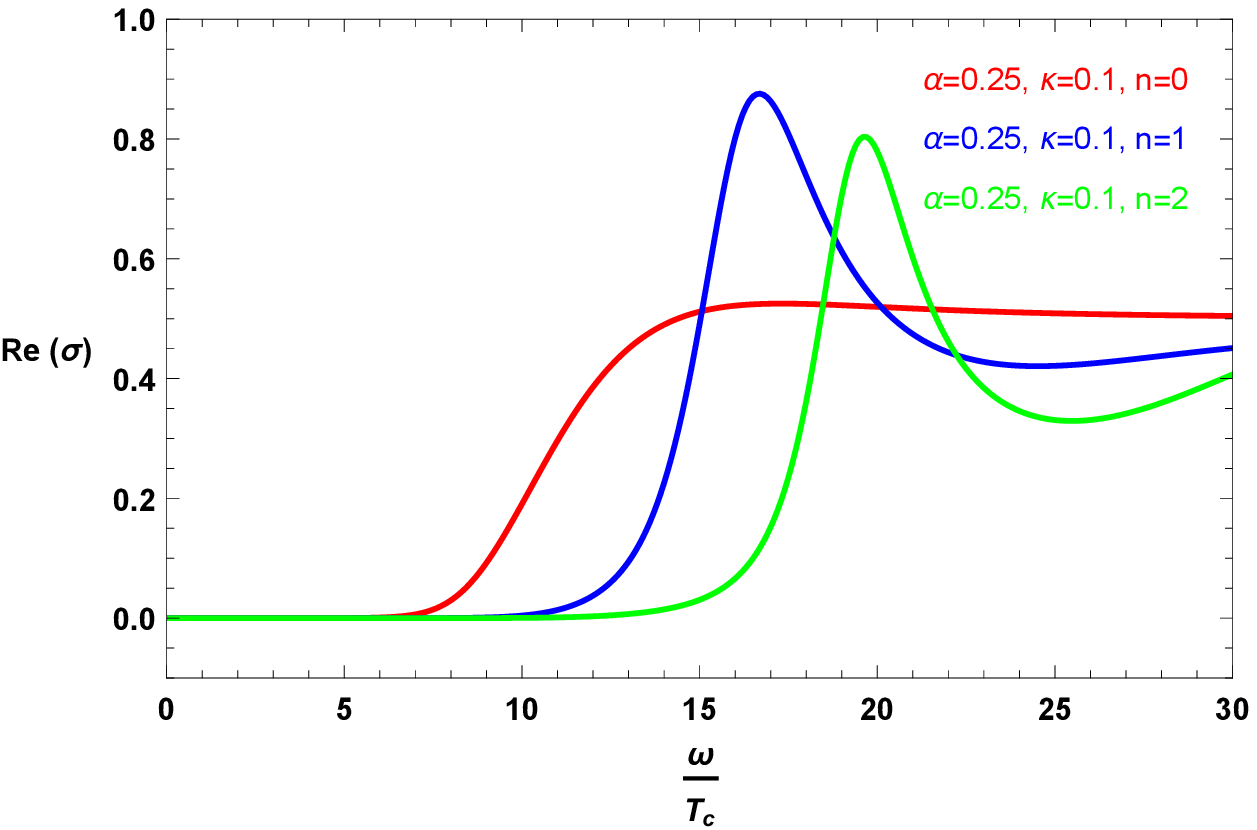}
\includegraphics[scale=0.43]{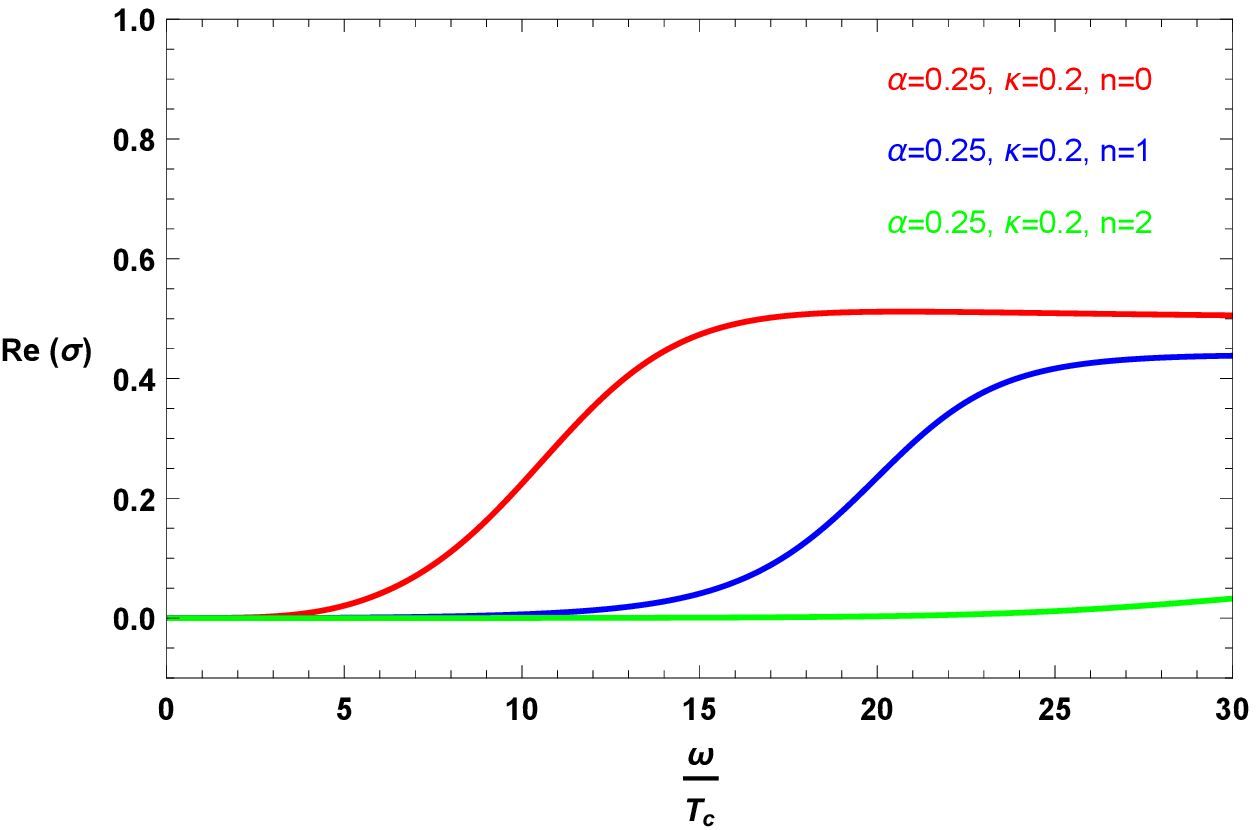}\\
\caption{\label{RealConductivity} (color online) Real part of the conductivity in the ($2+1$)-dimensional superconductors for the fixed mass of the scalar field $m^{2}L_{\rm eff}^{2}=-2$ with different Gauss-Bonnet parameters $\alpha$ and backreaction parameters $\kappa$ in the ground and excited states. In each panel, the red, blue and green lines denote the ground ($n=0$), first ($n=1$) and second ($n=2$) states, respectively.}
\end{figure}

\begin{figure}[ht]
\includegraphics[scale=0.43]{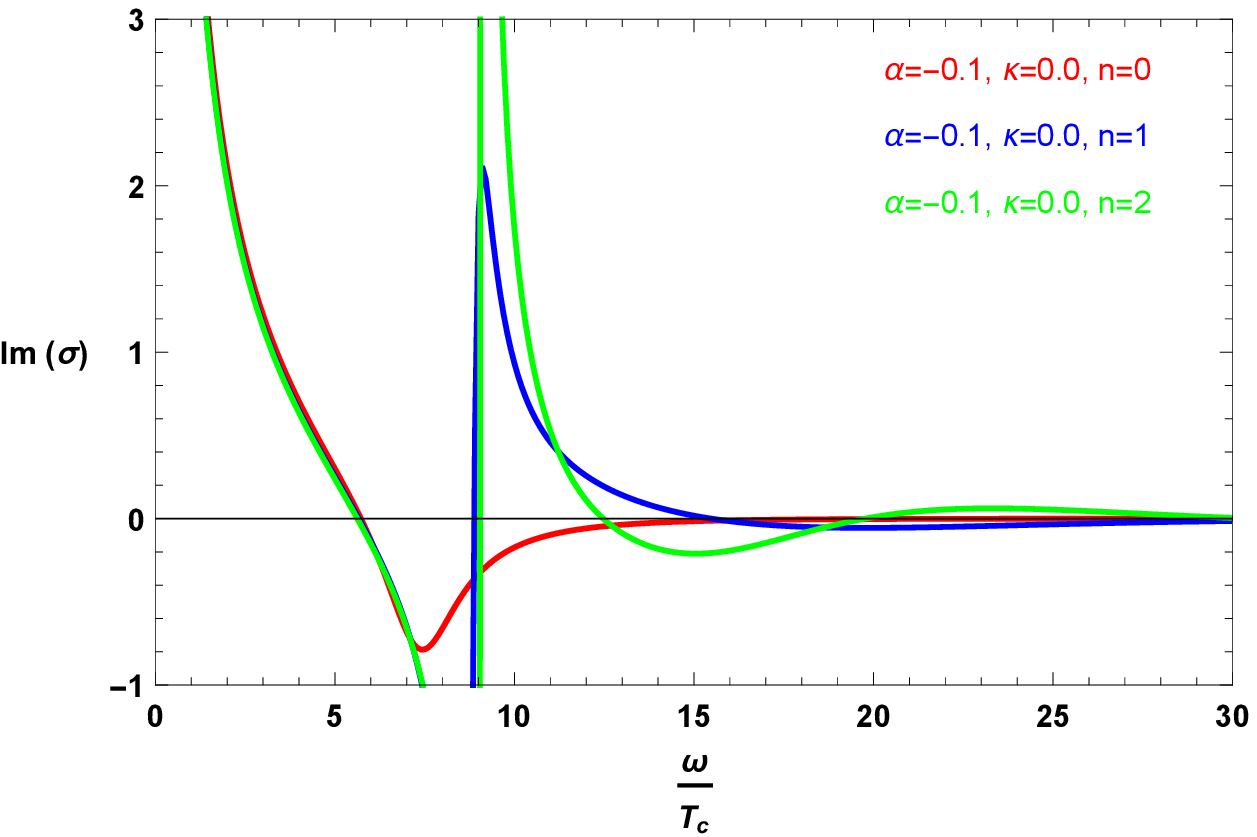}
\includegraphics[scale=0.43]{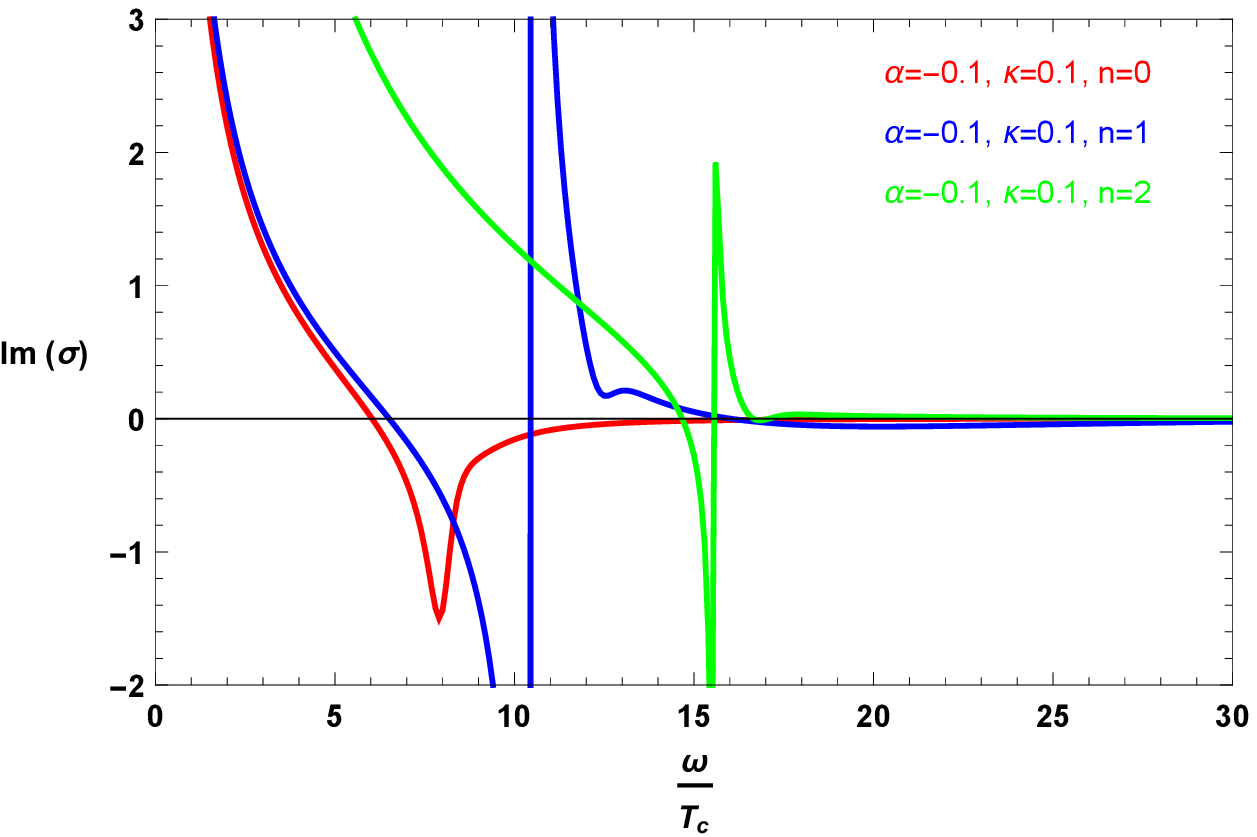}
\includegraphics[scale=0.43]{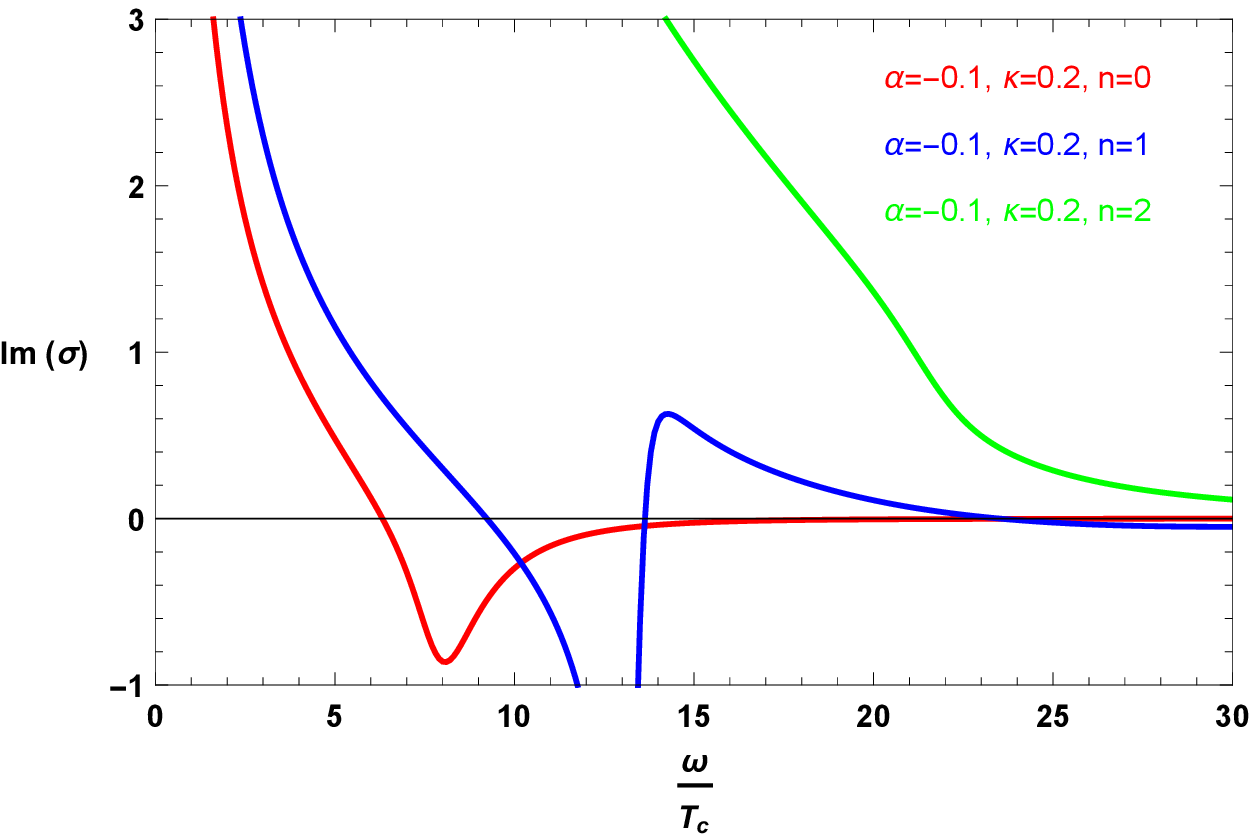}\\
\includegraphics[scale=0.43]{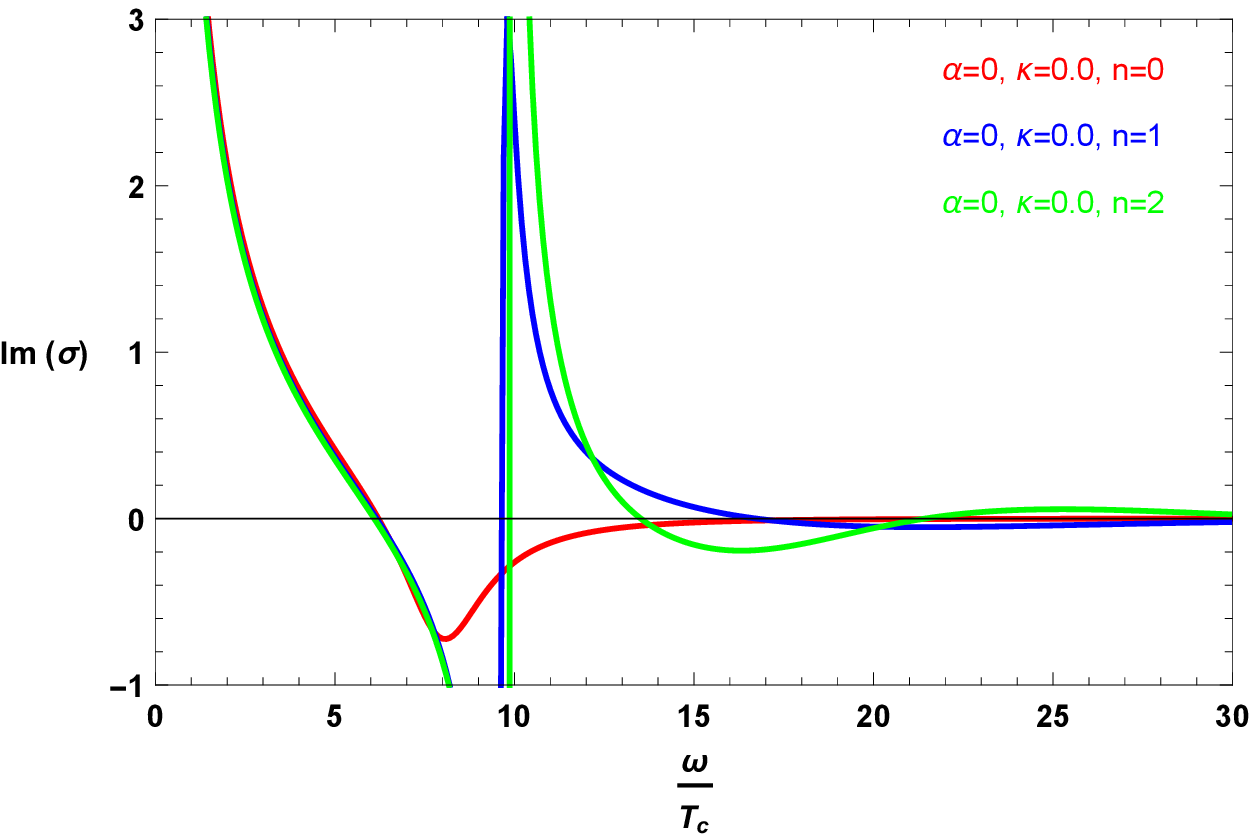}
\includegraphics[scale=0.43]{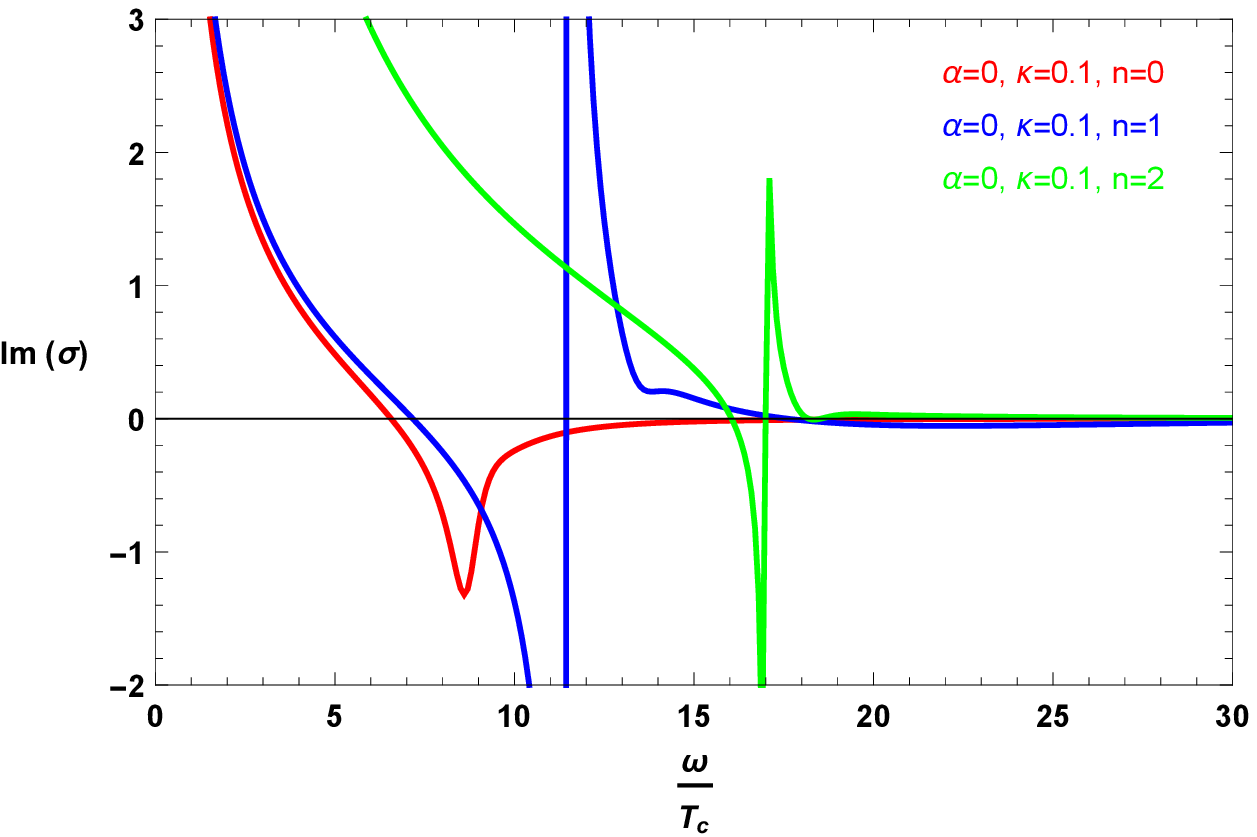}
\includegraphics[scale=0.43]{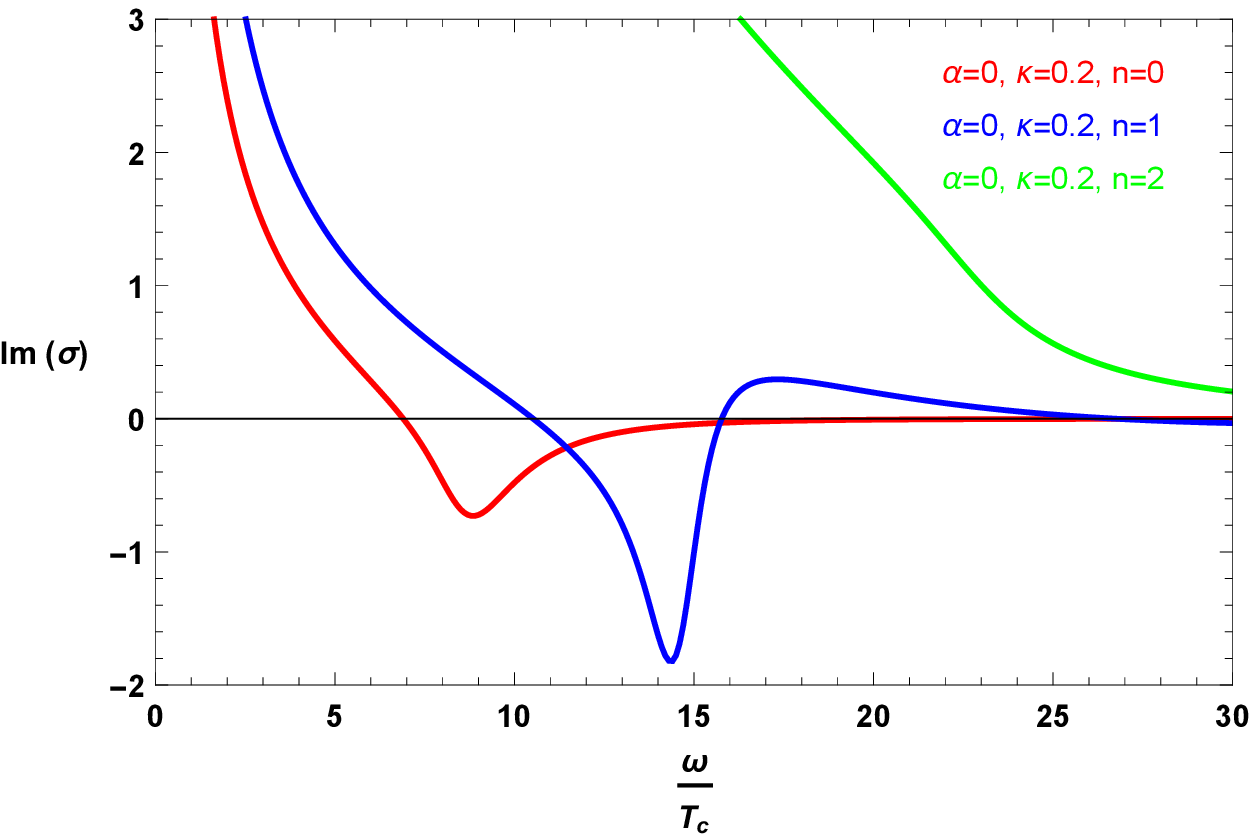}\\
\includegraphics[scale=0.43]{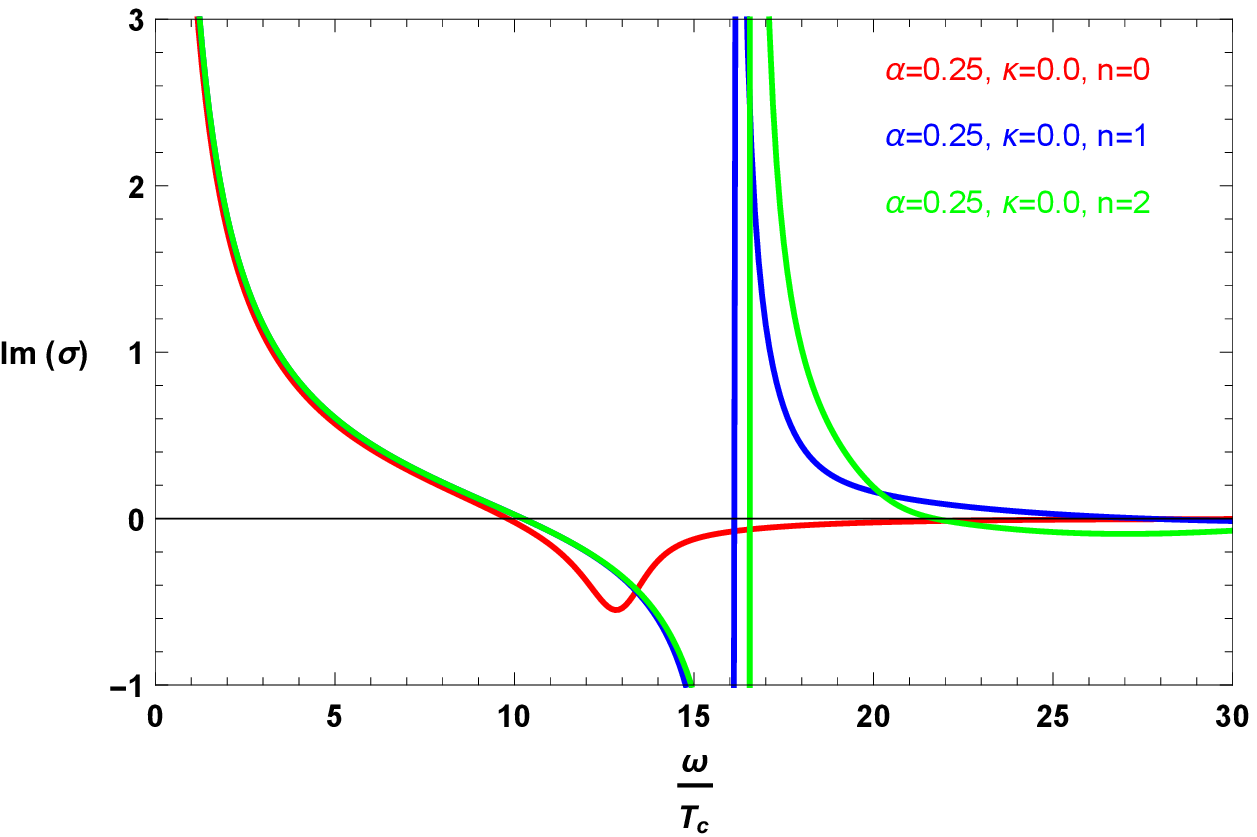}
\includegraphics[scale=0.43]{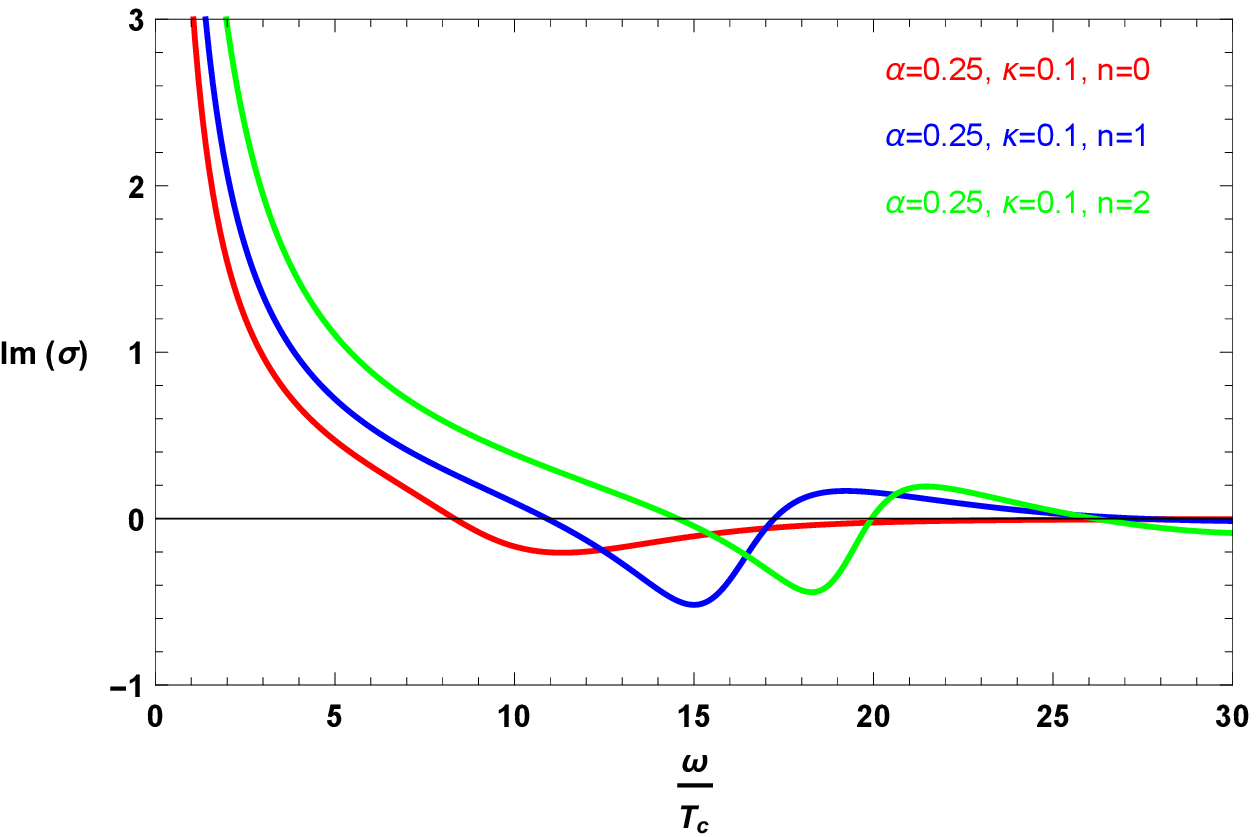}
\includegraphics[scale=0.43]{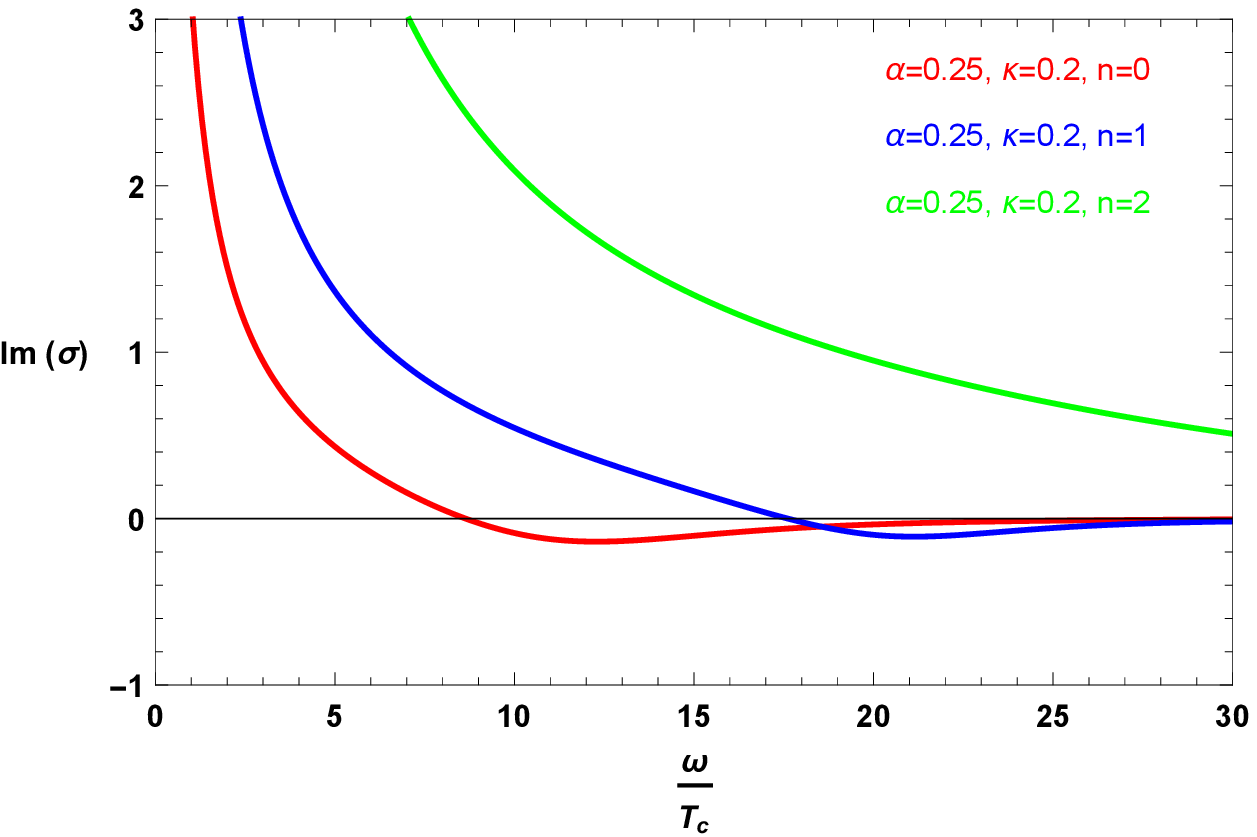}\\
\caption{\label{ImaginaryConductivity} (color online) Imaginary part of the conductivity in the ($2+1$)-dimensional superconductors for the fixed mass of the scalar field $m^{2}L_{\rm eff}^{2}=-2$ with different Gauss-Bonnet parameters $\alpha$ and backreaction parameters $\kappa$ in the ground and excited states. In each panel, the red, blue and green lines denote the ground ($n=0$), first ($n=1$) and second ($n=2$) states, respectively.}
\end{figure}

In order to study the excited states, in Figs. \ref{RealConductivity} and \ref{ImaginaryConductivity}, we present the frequency dependent conductivity $\sigma(\omega)$ of the ($2+1$)-dimensional superconductors for the scalar mass $m^{2}L_{\rm eff}^{2}=-2$ with different $\alpha$ and $\kappa$, where the red, blue and green lines denote the ground ($n=0$), first ($n=1$) and second ($n=2$) states, respectively. We can observe clearly that, there also exists a gap in the conductivity of the excited state and the gap frequency $\omega_g$ becomes larger when we increase the value of $\alpha$ or $\kappa$, which is similar to the ground state. It should be noted that for the excited states, there exist additional poles in Im[$\sigma$] and delta functions in Re[$\sigma$] arising at low temperature \cite{WangLLZEPJC}. Interestingly, we find that with the increase of $\alpha$ or $\kappa$, the pole and delta function can be broaden into the peaks with finite width. The combination of the Gauss-Bonnet gravity and backreaction provides richer physics in the conductivity $\sigma(\omega)$ of the ($2+1$)-dimensional superconductors.

\section{Conclusions}

We have constructed the ($2+1$)-dimensional superconductors beyond the probe limit in the consistent $D\rightarrow4$ Einstein-Gauss-Bonnet gravity proposed by Aoki, Gorji and Mukohyama \cite{AGM} in order to understand the influences of the $1/N$ or $1/\lambda$ corrections on the scalar condensate via the Abelian-Higgs model. Both for the ground state and excited states, we observed that the effect of the backreaction is to decrease the critical temperature $T_{c}$, which indicates that the backreaction can hinder the condensate to be formed. But, similar to that seen for the ($3+1$)-dimensional Gauss-Bonnet superconductors with the backreactions \cite{BarclayGregory}, the curvature correction has a more subtle effect: the critical temperature first decreases then increases as the curvature correction tends towards the Chern-Simons value in a backreaction dependent fashion, regardless of either the operator ${\cal O}_{+}$ or ${\cal O}_{-}$. Interestingly, the decrease of the backreaction $\kappa$, the increase of the scalar mass $m^2L_{\rm eff}^2$, or the increase of the number of nodes $n$ will weaken the subtle effect of the curvature correction on the critical temperature. Moreover, we studied the conductivity of the system, and found that, similar to the effect of the curvature correction for the higher-dimensional superconductors, the higher curvature corrections really alter the universal relation $\omega_g/T_c\approx 8$ for the ($2+1$)-dimensional superconductors with the backreactions. However, for the Gauss-Bonnet parameter $\alpha$ approaching the Chern-Simons limit, the gap frequency $\omega_g/T_c$ decreases first and then increases when $\kappa$ increases in a scalar mass dependent fashion, which is different from the finding in the ($3+1$)-dimensional superconductors that increasing backreaction parameter $\kappa$ increases $\omega_g/T_c$ in the full parameter space. This exhibits a very interesting and different feature when compared to the higher-dimensional Gauss-Bonnet superconductors. In addition, we noted that with the increase of the Gauss-Bonnet parameter $\alpha$ or backreaction $\kappa$, the pole and delta function of the conductivity for the ground state and excited states can be broaden into the peaks with finite width. Obviously, although how the curvature correction works in the holographic superconductors is still an open question, the combination of the Gauss-Bonnet gravity and backreaction provides richer physics in the condensates of the scalar hair and the conductivity in the ($2+1$)-dimensional superconductors.

{\bf Note added}------While we were completing this work, a complementary paper of D. Ghorai and S. Gangopadhyay \cite{GhoraiS2021} on the analytical holographic superconductor with the backreactions in the $4D$ Einstein-Gauss-Bonnet gravity appeared in arXiv.

\begin{acknowledgments}

This work was supported by the National Key Research and Development Program of China (No. 2020YFC2201400) and National Natural Science Foundation of China under Grant Nos. 11775076, 11965013, 12035005 and 11690034.

\end{acknowledgments}


\begin{thebibliography}{99}

\bibitem{Maldacena}
J. Maldacena,
%\textit{The large-N limit of superconformal field theories and supergravity},
Adv. Theor. Math. Phys. {\bf 2}, 231 (1998) [Int. J. Theor. Phys. {\bf 38}, 1113 (1999)].

\bibitem{Witten}
E. Witten,
%\textit{Anti-de Sitter space and holography},
Adv. Theor. Math. Phys. {\bf 2}, 253 (1998).

\bibitem{Gubser1998}
S.S. Gubser, I.R. Klebanov, and A.M. Polyakov,
%\textit{Gauge theory correlators from non-critical string theory},
Phys. Lett. B {\bf 428}, 105 (1998).

\bibitem{JZaanenSLS}
J. Zaanen, Y.W. Sun, Y. Liu, and K. Schalm, Holographic duality in
condensed matter physics (Cambridge, United Kingdom, Cambridge University Press, 2015).

\bibitem{GubserPRD78}
S.S. Gubser, Phys. Rev. D {\bf 78}, 065034 (2008).

\bibitem{HartnollPRL101}
S.A. Hartnoll, C.P. Herzog, and G.T. Horowitz,
%\textit{Building a Holographic Superconductor},
Phys. Rev. Lett. {\bf 101}, 031601 (2008).

\bibitem{HartnollJHEP12}
S.A. Hartnoll, C.P. Herzog, and G.T. Horowitz, %\textit{Holographic Superconductors},
J. High Energy Phys. {\bf 12}, 015 (2008).

\bibitem{FadafanRE}
K.B. Fadafan, J.C. Rojas, and N. Evans, %\textit{Holographic description of color superconductivity},
Phys. Rev. D {\bf 98}, 066010 (2018); arXiv:1803.03107 [hep-ph].


\bibitem{CaiRev}
R.G. Cai, L. Li, L.F. Li, and R.Q. Yang, %\textit{Introduction to Holographic Superconductor Models},
Sci. China-Phys. Mech. Astron. {\bf 58}, 060401 (2015); arXiv:1502.00437 [hep-th].

\bibitem{HartnollRev}
S.A. Hartnoll,
%\textit{Lectures on holographic methods for condensed matter physics},
Class. Quant. Grav. {\bf 26}, 224002 (2009).

\bibitem{HerzogRev}
C.P. Herzog, %\textit{Lectures on Holographic Superfluidity and Superconductivity},
J. Phys. A {\bf 42}, 343001 (2009).

\bibitem{HorowitzRev}
G.T. Horowitz, %\textit{Introduction to Holographic Superconductors},
Lect. Notes Phys. {\bf 828}, 313 (2011); arXiv:1002.1722 [hep-th].

\bibitem{Cai-2002}
R.G. Cai, %\textit{Gauss-Bonnet black holes in AdS spaces},
Phys. Rev. D {\bf 65}, 084014 (2002); arXiv:hep-th/0109133.

\bibitem{LiuFLWZ}
Y. Liu, G.Y. Fu, H.L. Li, J.P. Wu, and X. Zhang,
%\textit{Holographic p-wave superconductivity from higher derivative theory},
Eur. Phys. J. C {\bf 81}, 568 (2021); arXiv:2011.07330 [hep-th].

\bibitem{Gregory}
R. Gregory, S. Kanno, and J. Soda, J. High Energy Phys. {\bf 10},
010 (2009).

\bibitem{HorowitzPRD78}
G.T. Horowitz and M.M. Roberts, Phys. Rev. D {\bf 78}, 126008
(2008).

\bibitem{BarclayGregory}
L. Barclay, R. Gregory, S. Kanno, and P. Sutcliffe,
%\textit{Gauss-Bonnet Holographic Superconductors},
J. High Energy Phys. {\bf 12}, 029 (2010); arXiv: 1009.1991 [hep-th].

\bibitem{Brihaye}
Y. Brihaye and B. Hartmann, Phys. Rev. D {\bf 81}, 126008 (2010);
arXiv:1003.5130 [hep-th].

\bibitem{Pan-Wang}
Q.Y. Pan, B. Wang, E. Papantonopoulos, J. Oliveria, and A.B. Pavan,
Phys. Rev. D {\bf 81}, 106007 (2010).

\bibitem{Gregory2011}
R. Gregory,
%\textit{Holographic Superconductivity with Gauss-Bonnet gravity},
J. Phys. Conf. Ser. {\bf 283}, 012016 (2011); arXiv:1012.1558
[hep-th]

\bibitem{Ge-Wang}
X.H. Ge, B. Wang, S.F. Wu, and G.H. Yang,
%\textit{Analytical study on holographic superconductors in external magnetic field},
J. High Energy Phys. {\bf 08}, 108 (2010); arXiv:1002.4901 [hep-th].

\bibitem{KannoGB}
S. Kanno,
%\textit{A Note on Gauss-Bonnet Holographic Superconductors},
Class. Quant. Grav. {\bf 28}, 127001 (2011).

\bibitem{Gangopadhyay2012}
S. Gangopadhyay and D. Roychowdhury,
%\textit{Analytic study of Gauss-Bonnet holographic superconductors in Born-Infeld electrodynamics},
J. High Energy Phys. {\bf 05}, 156 (2012).

\bibitem{GhoraiGangopadhyay}
D. Ghorai and S. Gangopadhyay,
%\textit{Higher dimensional holographic superconductors in Born-Infeld electrodynamics with back-reaction},
Eur. Phys. J. C {\bf 76}, 146 (2016).

\bibitem{SheykhiSalahiMontakhab}
A. Sheykhi, H.R. Salahi, and A. Montakhab,
%\textit{Analytical and Numerical Study of Gauss-Bonnet Holographic Superconductors with Power-Maxwell Field},
J. High Energy Phys. {\bf 04}, 058 (2016).

\bibitem{SalahiSheykhiMontakhab}
H.R. Salahi, A. Sheykhi, and A. Montakhab,
%\textit{Effects of Backreaction on Power-Maxwell Holographic Superconductors in Gauss-Bonnet Gravity},
Eur. Phys. J. C {\bf 76}, 575 (2016); arXiv:1608.05025 [gr-qc].

\bibitem{LiFuNie}
Z.H. Li, Y.C. Fu, and Z.Y. Nie,
%Competing s-wave orders from Einstein-Gauss-Bonnet gravity
Phys. Lett. B {\bf 776}, 115 (2018).

\bibitem{CHNam}
C.H. Nam, Phys. Lett. B {\bf 797}, 134865 (2019) ; Gen. Rel. Grav. {\bf 51}, 104 (2019).

\bibitem{ParaiEPJC2020}
D. Parai, D. Ghorai, and S. Gangopadhyay,
%Effect of magnetic field on holographic insulator/superconductor phase transition in higher dimensional Gauss-Bonnet gravity
Eur. Phys. J. C {\bf 80}, 232 (2020).

\bibitem{CaiPWaveGB}
R.G. Cai, Z.Y. Nie, and H.Q. Zhang, %Holographic p-wave superconductors from Gauss-Bonnet gravity,
Phys. Rev. D {\bf 82}, 066007 (2010); arXiv:1007.3321 [hep-th].

\bibitem{LiCaiZhang}
H.F. Li, R.G. Cai, and H.Q. Zhang,
%\textit{Analytical Studies on Holographic Superconductors in Gauss-Bonnet Gravity},
J. High Energy Phys. {\bf 04}, 028 (2011).

\bibitem{LuWuNPB2016}
J.W. Lu, Y.B. Wu, T. Cai, H.M. Liu, Y.S. Ren, and M.L. Liu,
%Holographic vector superconductor in Gauss-Bonnet gravity
Nucl. Phys. B {\bf 903}, 360 (2016).

\bibitem{GBSuperfluid}
S.C. Liu, Q.Y. Pan, and J.L. Jing, Phys. Lett. B {\bf 765}, 91 (2017); arXiv:1610.02549 [hep-th].

\bibitem{MohammadiEPJC2019}
M. Mohammadi and A. Sheykhi,
%Conductivity of the holographic p-wave superconductors with higher order corrections
Eur. Phys. J. C {\bf 79}, 743 (2019).

\bibitem{LaiEPJC2020}
C.Y. Lai, T.M. He, Q.Y. Pan, and J.L. Jing,
%Analytical study of holographic p-wave superfluid models in Gauss-Bonnet gravity
Eur. Phys. J. C {\bf 80}, 247 (2020).

\bibitem{NieZeng}
Z.Y. Nie and H. Zeng
%\textit{P-T phase diagram of a holographic s+p model from Gauss-Bonnet gravity},
J. High Energy Phys. {\bf 10}, 047 (2015).

\bibitem{FadafanRojas}
K.B. Fadafan and J.C. Rojas
%\textit{Holographic Colour Superconductors at Finite Coupling with NJL Interactions}, 
arXiv:2107.04299 [hep-ph].

\bibitem{GlavanLin}
D. Glavan and C.S. Lin, Phys. Rev. Lett. {\bf 124}, 081301 (2020); arXiv:1905.03601 [gr-qc].

\bibitem{Ai2020}
W.Y. Ai,
%A note on the novel 4D Einstein-Gauss-Bonnet gravity,
Commun. Theor. Phys. {\bf 72}, 095402 (2020); arXiv:2004.02858 [gr-qc].

\bibitem{Mahapatra2020}
S. Mahapatra,
%A note on the total action of 4D Gauss-Bonnet theory,
Eur. Phys. J. C {\bf 80}, 992 (2020); arXiv:2004.09214 [gr-qc].

\bibitem{Shu09339}
F.W. Shu,
%Vacua in novel 4D Einstein-Gauss-Bonnet Gravity: pathology and instability?
Phys. Lett. B {\bf  811}, 135907 (2020); arXiv:2004.09339 [gr-qc].

\bibitem{TianZhu}
S.X. Tian and Z.H. Zhu,
%Comment on ``Einstein-Gauss-Bonnet Gravity in Four-Dimensional Spacetime"
arXiv:2004.09954 [gr-qc].

\bibitem{ArrecheaDJ}
J. Arrechea, A. Delhom, and A. Jim\'{e}nez-Cano,
%Yet another comment on four-dimensional Einstein-Gauss-Bonnet gravity
Chin. Phys. C {\bf 45}, 013107 (2021); arXiv:2004.12998 [gr-qc].

\bibitem{GursesST}
M. G\"{u}rses, T.\c{C}. \c{S}i\c{s}man, and B. Tekin,
%Is there a novel Einstein-Gauss-Bonnet theory in four dimensions?
Eur. Phys. J. C {\bf 80}, 647 (2020);
%Comment on ``Einstein-Gauss-Bonnet Gravity in Four-Dimensional Spacetime"
Phys. Rev. Lett. {\bf 125}, 149001 (2020).

\bibitem{LuPang}
H. Lu and Y. Pang,
%Horndeski Gravity as D¡ú4 Limit of Gauss-Bonnet
Phys. Lett. B {\bf 809}, 135717 (2020); arXiv:2003.11552 [gr-qc].

\bibitem{HennigarKMP}
R.A. Hennigar, D. Kubiznak, R.B. Mann, and C. Pollack,
%On Taking the D¡ú4 limit of Gauss-Bonnet Gravity: Theory and Solutions
J. High Energy Phys. {\bf 07}, 027 (2020); arXiv:2004.09472 [gr-qc].

\bibitem{Fernandes08362}
P.G.S. Fernandes, P. Carrilho, T. Clifton, and D.J. Mulryne,
%Derivation of Regularized Field Equations for the Einstein-Gauss-Bonnet Theory in Four Dimensions
Phys. Rev. D {\bf 102}, 024025 (2020); arXiv:2004.08362 [gr-qc].

\bibitem{OikonomouF}
V.K. Oikonomou and F.P. Fronimos,
%Reviving non-Minimal Horndeski-like Theories after GW170817: Kinetic Coupling Corrected Einstein-Gauss-Bonnet Inflation
Class. Quantum Grav. {\bf 38}, 035013 (2021); arXiv:2006.05512 [gr-qc].

\bibitem{AGM}
K. Aoki, M.A. Gorji, and S. Mukohyama,
%A consistent theory of D\rightarrow4 Einstein-Gauss-Bonnet gravity
Phys. Lett. B {\bf 810}, 135843 (2020); arXiv:2005.03859 [gr-qc].

\bibitem{qiao}
X.Y. Qiao, L. OuYang, D. Wang, Q.Y. Pan, and J.L. Jing,
%Holographic superconductors in 4D Einstein-Gauss-Bonnet gravity
J. High Energy Phys. {\bf 12}, 192 (2020); arXiv:2005.01007 [hep-th].

\bibitem{Fernandes}
P.G.S. Fernandes, %Charged Black Holes in AdS Spaces in 4D Einstein Gauss-Bonnet Gravity
Phys. Lett. B {\bf 805}, 135468 (2020); arXiv:2003.05491 [gr-qc].

\bibitem{WeiL14275}
S.W. Wei and Y.X. Liu,
%Extended thermodynamics and microstructures of four-dimensional charged Gauss-Bonnet black hole in AdS space
Phys. Rev. D {\bf 101}, 104018 (2020); arXiv:2003.14275 [gr-qc].

\bibitem{KonoplyaZhidenko}
R.A. Konoplya and A. Zhidenko,
%Black holes in the four-dimensional Einstein-Lovelock gravity
Phys. Rev. D {\bf 101}, 084038 (2020); arXiv:2003.07788 [gr-qc].

\bibitem{WangJHEP2020}
Y.Q. Wang, T.T. Hu, Y.X. Liu, J. Yang, and L. Zhao,
%\textit{Excited states of holographic superconductors},
J. High Energy Phys. {\bf 06}, 013 (2020); arXiv:1910.07734 [hep-th].

\bibitem{QiaoEHS}
X.Y. Qiao, D. Wang, L. OuYang, M.J. Wang, Q.Y. Pan, and J.L. Jing,
%\textit{An analytic study on the excited states of holographic superconductors},
Phys. Lett. B {\bf 811}, 135864 (2020); arXiv:2007.08857 [hep-th].

\bibitem{Liran}
R. Li, J. Wang, Y. Q. Wang, and H.B. Zhang,
%\textit{\Nonequilibrium dynamical transition process betweenexcited states of holographic superconductors},
J. High Energy Phys. {\bf 11}, 059 (2020); arXiv:2008.07311 [hep-th].

\bibitem{XiangZW}
Q. Xiang, L. Zhao, and Y.Q. Wang,
%\textit{Excited states of holographic superconductors from massive gravity},
arXiv:2010.03443 [hep-th].

\bibitem{OuYangliang}
L. Ouyang, D. Wang, X.Y. Qiao, M.J. Wang, Q.Y. Pan, and J.L. Jing,
%\textit{Holographic insulator/superconductor phase transitions with excited states},
Sci. China Phys. Mech. Astron. {\bf 64}, 240411 (2021); arXiv:2010.10715 [hep-th].

\bibitem{WangLLZEPJC}
Y.Q. Wang, H.B. Li, Y.X. Liu, and Y. Zhong,
%\textit{Excited states of holographic superconductors with backreaction},
 Eur. Phys. J. C {\bf 81}, 628 (2021); arXiv:1911.04475 [hep-th].

\bibitem{ZhangZPJ}
S.H. Zhang, Z.X. Zhao, Q.Y. Pan, and J.L. Jing,
%Excited states of holographic superconductors with hyperscaling violation
arXiv:2107.09486 [hep-th].

\bibitem{AokiGMJCAP}
K. Aoki, M.A. Gorji, and S. Mukohyama,
%Cosmology and gravitational waves in consistent D\rightarrow4 Einstein-Gauss-Bonnet gravity
J. Cosmol. Astropart. Phys. {\bf 09}, 014 (2020); arXiv:2005.08428 [gr-qc].

\bibitem{AokiGMM}
K. Aoki, M.A. Gorji, S. Mizuno, and S. Mukohyama,
%Inflationary gravitational waves in consistent D¡ú4 Einstein-Gauss-Bonnet gravity
J. Cosmol. Astropart. Phys. {\bf 01}, 054 (2021); arXiv:2010.03973 [gr-qc].

\bibitem{PanJWC}
Q.Y. Pan, J.L. Jing, B. Wang, and S.B. Chen,
%\textit{Analytical study on holographic superconductors with backreactions},
J. High Energy Phys. {\bf 06}, 087 (2012); arXiv:1205.3543 [hep-th].

\bibitem{CrisostomoTZ}
J. Crisostomo, R. Troncoso, and J. Zanelli, Phys. Rev. D {\bf 62}, 084013 (2000).

\bibitem{WangSPJ}
D. Wang, M.M. Sun, Q.Y. Pan, and J.L. Jing,
%\textit{Backreacting holographic superconductors from the coupling of a scalar field to the Einstein tensor},
Phys. Lett. B {\bf 785}, 362 (2018).

\bibitem{GhoraiS2021}
D. Ghorai and S. Gangopadhyay,
%Analytical study of holographic superconductor with backreaction in 4d Gauss-Bonnet gravity
arXiv:2105.09423 [hep-th].


\end{thebibliography}
\end{document}